\documentclass[11pt]{article}
\usepackage[margin=0.7in]{geometry}
\usepackage{amsmath,amssymb}
\usepackage{booktabs,array}
\usepackage{graphicx}
\usepackage[flushleft]{threeparttable}
\usepackage{setspace}
\usepackage{authblk}
\usepackage[colorlinks=true,linkcolor=blue,citecolor=blue]{hyperref}
\usepackage{float}
\title{Bayesian Recovery Trajectories Following Forced Sexual Experience: Healthcare Access and Social Support Among Women}

\author[1*]{Hammed A. Olayinka}
\author[2]{Saheed O. Olayinka}
\affil[1]{Department of Mathematical Sciences, Worcester Polytechnic Institute, Worcester, MA, USA}
\affil[2]{Department of Computer Science and Artificial Intelligence, University of Ibadan, Ibadan, Nigeria}
\affil[*]{\texttt{haolayinka@wpi.edu}}
\date{}

\begin{document}
\maketitle

\begin{abstract}
\noindent\textbf{Background.} Women who experience forced sexual violence are at elevated risk of poor mental and physical health, but whether recovery follows a single course or several distinct trajectories, and whether healthcare access and social support distinguish them, is unclear in nationally representative, multi-decade cohorts.

\noindent\textbf{Methods.} We used data from 1,848 women in the National Longitudinal Study of Adolescent to Adult Health (Add Health), followed from adolescence into adulthood. An identified Bayesian mixture model classified women into joint mental and physical health burden trajectories from six repeated indicators. Bayesian logistic models estimated associations between forced sexual experience, healthcare access, social support, and demographics and trajectory membership. Robustness was evaluated across three mixture specifications and against four- and five-class alternatives.

\noindent\textbf{Results.} A three-class solution was retained: low-burden/resilient (27.2\% posterior mean class weight), moderate/worsening burden (43.2\%), and high chronic burden (29.7\%); all specifications converged cleanly ($\hat{R}=1.00$) with clearly separated classes, whereas four- and five-class alternatives showed convergence problems or an unstable additional class. Forced sexual experience was associated with higher posterior odds of high chronic burden membership (OR = 3.78, 95\% credible interval [CrI] 2.84--5.04) and moderate/worsening membership (OR = 1.63, 95\% CrI 1.26--2.12). Better healthcare access and greater social connectedness were independently associated with lower posterior odds of high chronic burden membership.

\noindent\textbf{Conclusions.} Long-term health after forced sexual experience is heterogeneous, not uniform, and robust to modeling choices. Healthcare access and social connectedness distinguish more from less favorable trajectories, identifying two modifiable targets for trauma-informed care. These findings highlight the value of modeling heterogeneous recovery patterns and identify healthcare access and social connectedness as modifiable targets for trauma-informed, longitudinal, and survivor-centered care.
\end{abstract}


\section*{Introduction}

Sexual violence is a major public health problem with consequences that may persist across the life course. Women who experience sexual violence are at increased risk for posttraumatic stress, depression, anxiety, psychological distress, suicidality, chronic pain, sleep difficulty, functional limitations, and poor self-rated health (Dworkin et al., 2017; Basile et al., 2021; Basile et al., 2024). This pattern of co-occurring psychological and physical health consequences echoes the broader literature on intimate partner violence against women (Campbell, 2008; Bacchus et al., 2018; White et al., 2023), and national and global surveillance data document the substantial scope of the problem (Garcia-Moreno et al., 2006; Smith et al., 2018; WHO, 2021; Olalude et al., 2024), underscoring the need for longitudinal, survivor-centered research that examines long-term well-being rather than only acute harm. Much of the existing evidence on health consequences of sexual violence comes from cross-sectional or short-follow-up studies, which can document that mental and physical health are, on average, worse among women with a history of forced sexual experience, but cannot characterize whether this average difference reflects a uniform, moderate elevation in risk shared by most survivors or, alternatively, a more polarized pattern in which most women recover while a subset experiences persistently severe and worsening health.

Most existing longitudinal work on recovery after sexual violence has two important limitations. First, mental and physical health are typically modeled separately, even though both are commonly affected and may covary in ways that a single-domain model cannot capture; a woman whose depressive symptoms improve but whose physical health continues to decline, for example, would be classified as recovering by a mental-health-only model despite an unfavorable overall health course. Second, prior longitudinal studies have generally relied on clinical, convenience, or help-seeking samples recruited shortly after an index assault (Steenkamp et al., 2012; Goodman-Williams et al., 2024; Campbell et al., 2011; Short et al., 2019, 2020), which limits generalizability to the broader population of women with a lifetime history of forced sexual experience -- many of whom never present to clinical or forensic services -- and precludes multi-decade follow-up into adulthood.

The broader trauma literature has identified common posttraumatic trajectories -- including resilience, recovery, chronic distress, and delayed deterioration -- rather than a single average course following adversity (Bonanno, 2004; Bonanno et al., 2011; Galatzer-Levy et al., 2018; Norris et al., 2009; Pietrzak et al., 2014). Whether this heterogeneity extends to the \emph{joint} course of mental and physical health specifically after forced sexual experience, in a population-based cohort followed from adolescence into adulthood, has not been established. Identifying such heterogeneity has direct clinical relevance: if most women follow a single, moderately unfavorable course, universal early intervention may be the more efficient strategy, whereas if a subset of women follows a persistently high-burden course, targeted, sustained intervention for that subgroup may be more appropriate. We also examine whether two modifiable, post-assault resources -- healthcare access and social connectedness -- are associated with more favorable trajectory membership, since both represent plausible, actionable targets for trauma-informed clinical and public health intervention, in contrast to non-modifiable exposure history itself.

This study makes four contributions relative to this literature. First, it jointly models mental and physical health recovery, rather than a single symptom domain, among women with forced sexual experience. Second, it draws on a nationally representative, multi-decade adult cohort rather than a clinical or help-seeking sample recruited shortly after an index assault. Third, it applies identified Bayesian mixture estimation, explicitly addressing label-switching and propagating parameter and classification uncertainty into every downstream estimate, and directly tests the robustness of the resulting class structure against alternative numbers of classes. Fourth, it tests whether healthcare access and social connectedness distinguish trajectory classes, moving beyond documentation of average harm toward mechanisms that trauma-informed care and policy could plausibly act on.

This work is also motivated by a life-course health development perspective, which conceptualizes health as an emergent property of cumulative biological, behavioral, and social exposures across development rather than a fixed trait (Halfon et al., 2014). Under this framework, forced sexual experience during adolescence or young adulthood is not simply a discrete event with a bounded aftermath, but a potential inflection point that can alter the subsequent developmental trajectory of both mental and physical health, interacting with ongoing stress processes (Pearlin et al., 1981) and with the resources, such as healthcare access and social support, that are available to buffer or redirect that trajectory. A life-course perspective therefore predicts precisely the kind of heterogeneity this study investigates: because women differ in the resources available to them after an assault, they need not converge on a single average health trajectory, and identifying the resources associated with more favorable trajectories is a natural extension of this framework into actionable intervention targets.

\section*{Methods}

\subsection*{Data source and sample}

We used the National Longitudinal Study of Adolescent to Adult Health (Add Health), a nationally representative cohort that began with a school-based sample of adolescents in grades 7--12 in 1994--1995 and has followed respondents through six waves of data collection into adulthood (Griffiths et al., 2026). Add Health collects repeated information on mental and physical health, healthcare access and utilization, family and peer relationships, and socioeconomic circumstances, using in-home interviews and self-administered questionnaires.

The present analysis began with 2,629 female Add Health respondents with valid Wave III data. The analytic sample was restricted to those with complete data on six repeated joint trajectory indicators (depressive symptom burden and self-rated physical health burden at Waves III, IV, and V): 296 women were excluded for missing Wave IV depressive symptom data, a further 483 for missing Wave V depressive symptom data, and a further 2 for missing Wave V physical health data, yielding a final analytic sample of 1,848 women. A demographic-adjusted class-membership subsample additionally required complete covariate data (Wave IV healthcare access, Wave IV closeness to friends, Hispanic ethnicity, and race), excluding a further 43 women (9 for missing Wave IV healthcare access, 21 for missing Wave IV closeness to friends, and 13 for missing race or ethnicity data), yielding 1,805 women for the fully adjusted class-membership model.

Women excluded from the analytic sample because of incomplete trajectory data (n = 781) differed from those retained on several baseline characteristics: excluded women were slightly older (mean age 21.93 vs.\ 21.64 years, $p<.001$), less likely to report forced sexual experience at the baseline exposure items (23.1\% vs.\ 33.3\%, $p<.001$), more likely to identify as Hispanic (13.9\% vs.\ 9.1\%, $p<.001$) or Black or African American (30.3\% vs.\ 24.4\%, $p=.002$), and less likely to identify as White (62.2\% vs.\ 71.8\%, $p<.001$); Wave III healthcare access did not differ meaningfully between included and excluded women (0.79 vs.\ 0.81, $p=.33$). These differences indicate that attrition across Waves III--V was not random with respect to race, ethnicity, and forced sexual experience, and they are addressed further in the Limitations.

\subsection*{Exposure}

The primary exposure was a binary indicator of any reported forced sexual experience (physical or nonphysical), constructed from direct Wave IV and Wave V items, consistent with uniform surveillance definitions of sexual violence victimization (Basile et al., 2014). Respondents were coded as exposed if any direct Wave IV or Wave V forced-sex item indicated exposure, and as unexposed if valid information was available on these items and none indicated exposure. Physical and nonphysical forced sexual experience were also examined separately in sensitivity analyses to evaluate whether the primary combined exposure obscured differences between these two subtypes. Relationship-specific partner-forced-sex items, which would have allowed the primary exposure to incorporate a relationship context, were evaluated separately in a complete-case sensitivity analysis (Supplementary Material) rather than the primary definition, because response coding for these items required additional verification and because missingness was substantially higher than for the direct forced-sex items.

\subsection*{Trajectory indicators and candidate pathway variables}

Depressive symptom burden was the mean of four depressive symptomatology items at Wave III, five items at Wave IV, and four items at Wave V (internal consistency: Cronbach's $\alpha = 0.83$, 0.83, and 0.84, respectively), drawn from Add Health's standard depressive-symptom item battery at each wave. Self-rated physical health burden was a single self-rated general health item at each of the three waves. Both indicators were standardized prior to analysis so that the six trajectory indicators were on a common scale for mixture modeling; higher values on both indicators reflect greater burden (more depressive symptoms, or worse self-rated physical health). Because self-rated physical health was assessed with a single item, it should be interpreted as a broad indicator of physical health burden rather than a multidimensional physical health index. Candidate pathway variables, selected a priori as modifiable post-assault resources plausibly related to trajectory membership, were a healthcare access index at Waves III and IV (capturing whether respondents had a usual source of care and had seen a healthcare provider in the recall period) and a measure of closeness to friends at Wave IV.

\subsection*{Statistical analysis}

Latent joint trajectory classes were first enumerated using Gaussian mixture models (two to five classes) fit to the six standardized indicators, compared using BIC, AIC, silhouette values, and class size, consistent with standard growth-mixture class-enumeration practice (Muthén, 2002; Muthén \& Asparouhov, 2009). This preliminary stage was used only to inform priors for a subsequent, fully Bayesian mixture model, not as the final trajectory model; retaining it as a staging step, rather than treating its output as final, allowed the identified Bayesian model's priors to be genuinely informative without fixing any parameter to the preliminary estimates.

The primary Bayesian mixture model, estimated in PyMC 5.28.5 with four Markov chain Monte Carlo (MCMC) chains, used priors weakly centered on the preliminary Gaussian-mixture class proportions, means, and standard deviations:
\begin{align*}
\boldsymbol{\pi} &\sim \text{Dirichlet}(\alpha_k),\quad \alpha_k=\max(40\hat p_k,1.5); \qquad \mu_{kj}\sim N(\hat\mu_{kj},0.30^2);\\
\sigma_{kj} &\sim \text{TruncatedNormal}\!\left(\max(\hat\sigma_{kj},0.45),\,0.20^2;\,0.30,\,3.00\right),
\end{align*}
with an ordered class-severity constraint, $S_1<S_2<S_3$ where $S_k=\tfrac{1}{6}\sum_j \mu_{kj}$, imposed as a hard log-posterior penalty. This constraint resolves the label-switching problem inherent to finite mixture models, in which the likelihood is invariant to permutations of class labels, by ensuring that class 1 always denotes the lowest-severity class and class 3 the highest across all posterior draws. Model fit was additionally assessed by comparing observed, standardized class-specific indicator means against posterior model-implied class means for the assigned classes; small discrepancies would indicate that the estimated classes closely reproduce the observed data patterns they are intended to summarize, whereas large discrepancies would suggest model misspecification.

Robustness was assessed against a weakly identified specification (priors centered only on the rank order of severity, $\mu_{kj}\sim N(c_k,1.25^2)$ with fixed anchors $c=(-0.45,0,0.45)$, not on data-derived class-specific values) and an unconstrained specification (fully symmetric priors, $\mu_{kj}\sim N(0,2^2)$, no ordering constraint during sampling, posterior draws relabeled post hoc by severity), and against four- and five-class alternatives using an identical specification to the primary three-class model. The weakly identified and unconstrained specifications serve different robustness purposes: the former tests sensitivity to the informativeness of the priors while retaining the ordering constraint, and the latter tests whether the three-class structure emerges even without any constraint imposed during sampling, addressing the concern that an imposed ordering constraint could itself manufacture apparent class separation. Missingness on trajectory indicators was handled by complete-case restriction; covariate missingness in the class-membership models was likewise handled by complete-case restriction within each model. Early adversity indicators were not included in the adjusted models because they were missing for approximately 97--98\% of the trajectory sample.

Bayesian logistic class-membership models estimated posterior odds ratios (ORs) for membership in each less-favorable class relative to the low-burden/resilient class, with $N(0,2^2)$ priors on log-odds coefficients. Four nested specifications were estimated to evaluate confounding and mediation-adjacent structure: Model 1 included forced sexual experience and age; Model 2 added Wave III and Wave IV healthcare access; Model 3 added Wave IV closeness to friends; and Model 4, the fully adjusted model reported here, additionally included Hispanic ethnicity and race indicators. This progression allowed us to observe whether the exposure--class-membership association attenuated as candidate pathway variables and demographic covariates were introduced, which would be consistent with (though not proof of) a mediating or confounding role for those variables; in practice, the exposure association remained stable in magnitude across all four models. An exploratory equity analysis compared predicted class probabilities by Hispanic ethnicity using Model 4; further subgroup analyses (e.g., by insurance status, socioeconomic status, or education) were not pursued given sample size constraints in the public-use data. Convergence was assessed via the potential scale reduction statistic ($\hat{R}$) and effective sample size (ESS) for all reported parameters, with $\hat{R}\leq1.01$ and adequate ESS required for a specification to be considered adequately converged; all posterior summaries are reported as posterior medians (for odds ratios) or means (for mixture model parameters) with 95\% highest-density credible intervals.

This study was not preregistered, and its analyses fall into two tiers rather than a single confirmatory test. The primary, confirmatory analyses are the identified Bayesian mixture model, its two robustness specifications, the class-count sensitivity comparison, and the fully adjusted class-membership model (Model 4); these were the analyses used to answer the study's central questions and are interpreted with standard weight. All remaining analyses -- the nested Model 1--3 progression, the pathway and moderation analyses, the domain-specific and cross-domain models, the survey-weighted and alternative-exposure-definition sensitivity analyses, and the equity analysis -- are explicitly exploratory, are described as such throughout the Results and Discussion, and are not intended to support confirmatory claims. Given the number of exploratory analyses conducted, some associations may reflect chance; we did not apply a formal multiplicity correction, consistent with common practice in Bayesian analysis, where 95\% credible intervals already convey the substantial uncertainty in each individual exploratory estimate, but we accordingly avoid emphasizing any single exploratory p-value-adjacent threshold and instead interpret exploratory findings as hypothesis-generating. The sensitivity of the primary class-membership odds ratios to the choice of prior standard deviation on the logistic coefficients (evaluated at $N(0,1^2)$, $N(0,2^2)$, and $N(0,5^2)$) is reported in Supplementary Material; the direction and approximate magnitude of the key exposure association were stable across this range. This report follows the STROBE (Strengthening the Reporting of Observational Studies in Epidemiology) guidelines for cohort studies; a completed STROBE checklist is provided in Supplementary Material.

\section*{Results}

Forced sexual experience was reported by 33.3\% of the analytic sample (physical: 19.8\%; nonphysical: 29.8\%; Table~1). Women reporting forced sexual experience differed modestly from those who did not on several sample characteristics, including higher mean depressive symptom and physical health burden at Waves III and V and somewhat lower Wave III healthcare access (Table~1). Gaussian mixture enumeration showed continued improvement in BIC/AIC through five classes, but a three-class Bayesian mixture solution was retained (Figure~1): low-burden/resilient (posterior mean class weight 27.2\%, 95\% CrI 24.5--30.1\%), moderate/worsening burden (43.2\%, 95\% CrI 39.4--47.2\%), and high chronic burden (29.7\%, 95\% CrI 26.1--33.0\%), with clearly ordered severity ($\hat{R}=1.00$ for all key parameters across all three mixture specifications). By contrast, a four-class alternative showed degraded convergence (maximum $\hat{R}=1.07$, minimum ESS 42), driven by two classes with severity estimates only 0.016 apart; a five-class alternative converged cleanly but isolated a class representing only 2.7\% of the sample with a severity estimate only 0.066 from the highest-burden class. Both patterns indicate that the additional classes fragmented existing structure or captured an idiosyncratic subgroup rather than a generalizable trajectory (full detail: Supplementary Material).

Using final class assignments from the identified Bayesian mixture model, forced sexual experience remained consistently associated with higher posterior odds of less favorable class membership across all four nested class-membership models. In the unadjusted Model 1, forced sexual experience was associated with higher posterior odds of moderate/worsening burden membership (OR = 1.55, 95\% CrI 1.20--2.01) and, more strongly, high chronic burden membership (OR = 3.34, 95\% CrI 2.58--4.41). After adding Wave III and Wave IV healthcare access in Model 2, the association remained stable (moderate/worsening: OR = 1.55, 95\% CrI 1.20--2.01; high chronic: OR = 3.37, 95\% CrI 2.56--4.42), and better Wave III healthcare access was independently associated with lower posterior odds of both less-favorable classes (moderate/worsening: OR = 0.84, 95\% CrI 0.75--0.95; high chronic: OR = 0.65, 95\% CrI 0.56--0.75); Wave IV healthcare access was not clearly associated with either comparison. Adding Wave IV closeness to friends in Model 3 again left the exposure association essentially unchanged (moderate/worsening: OR = 1.55, 95\% CrI 1.20--2.02; high chronic: OR = 3.41, 95\% CrI 2.58--4.50), with more close friends independently associated with lower posterior odds of both classes (moderate/worsening: OR = 0.86, 95\% CrI 0.77--0.96; high chronic: OR = 0.66, 95\% CrI 0.57--0.75). That the exposure association neither attenuated nor amplified as healthcare access, social connectedness, and demographic covariates were successively introduced indicates that these variables are not acting as confounders of the exposure--trajectory association, consistent with their conceptualization as independent, potentially modifiable correlates of trajectory membership rather than as explanations for the exposure effect itself.

In the fully adjusted Model 4 (Table~2, Figure~2), forced sexual experience was associated with substantially higher posterior odds of high chronic burden membership relative to low-burden/resilient membership (OR = 3.78, 95\% CrI 2.84--5.04) and, to a lesser extent, moderate/worsening membership (OR = 1.63, 95\% CrI 1.26--2.12). Better Wave III healthcare access was associated with lower posterior odds of both less-favorable classes (high chronic: OR = 0.65, 95\% CrI 0.56--0.76; moderate/worsening: OR = 0.85, 95\% CrI 0.75--0.96), and greater Wave IV closeness to friends was similarly protective (high chronic: OR = 0.70, 95\% CrI 0.60--0.80; moderate/worsening: OR = 0.88, 95\% CrI 0.78--0.98). Hispanic ethnicity was associated with higher posterior odds of high chronic burden membership (OR = 1.72, 95\% CrI 1.07--2.82), and identifying as Black or African American was similarly associated with higher posterior odds of high chronic burden membership (OR = 2.38, 95\% CrI 1.08--5.47), although these demographic findings should be interpreted cautiously given the exploratory nature of the equity analysis and the wide credible intervals.

Posterior predicted probabilities (Figure~3) showed a corresponding shift in trajectory membership by exposure status: at the average covariate profile, the predicted probability of high chronic burden membership was 0.212 (95\% CrI 0.183--0.243) among women without reported forced sexual experience versus 0.410 (95\% CrI 0.343--0.478) among women reporting forced sexual experience, with a corresponding decrease in predicted low-burden/resilient membership (0.333 vs.\ 0.189). Posterior classification quality was good overall (mean maximum posterior class probability 0.865, median 0.931; 72.2\% of respondents classified with probability $\geq$0.80, 56.8\% with probability $\geq$0.90), and was highest for the high chronic burden class (mean 0.891) relative to the moderate/worsening (0.854) and low-burden/resilient (0.856) classes. Observed-versus-model-implied class profile checks supported adequate model fit: mean absolute differences between observed and posterior model-implied standardized class means were small across all three classes (0.008, 0.015, and 0.052 for the low-burden/resilient, moderate/worsening, and high chronic burden classes, respectively; maximum absolute differences 0.019, 0.028, and 0.069), indicating that the estimated classes closely reproduced the joint trajectory patterns observed in the data.

Exploratory pathway analyses evaluated healthcare access and social connectedness as candidate resources linking forced sexual experience to trajectory membership, using two complementary approaches rather than formal causal mediation. First, Bayesian resource models estimated the posterior association between forced sexual experience and each candidate pathway variable directly; forced sexual experience was associated with lower posterior healthcare access and social connectedness scores at the waves examined, consistent with the hypothesis that these resources could plausibly lie on a pathway between exposure and trajectory membership, although the cross-sectional and observational nature of these associations precludes causal interpretation. Second, among women reporting forced sexual experience, higher Wave III healthcare access was associated with a 9.6-percentage-point higher predicted probability of low-burden/resilient membership and a 10.4-percentage-point lower predicted probability of high chronic burden membership (one SD above versus below the mean, holding other covariates at sample means); higher Wave IV closeness to friends showed a similar pattern (high chronic burden: $-$15.2 percentage points). Moderation analyses testing whether healthcare access or social connectedness altered the exposure--class-membership association directly found no clear evidence of interaction (forced sexual experience $\times$ Wave III healthcare access, high chronic burden: OR = 0.75, 95\% CrI 0.53--1.03), indicating that these resources are associated with trajectory membership broadly rather than specifically buffering the effect of exposure.

Results were robust across a series of sensitivity analyses (Supplementary Material). Physical and nonphysical forced sexual experience showed similar, independently elevated associations with high chronic burden membership (OR = 4.57, 95\% CrI 3.23--6.58, and OR = 3.01, 95\% CrI 2.27--3.99, respectively), indicating that both forms of forced sexual experience carry substantial long-term health consequences rather than the association being driven by one subtype. Domain-specific Bayesian mixture models, estimated separately for mental and physical health indicators, each recovered a clearly ordered three-class structure with $\hat{R}=1.00$ for all key parameters (mental-health-only model: low mental burden 27.5\%, moderate 48.0\%, high 24.5\%; physical-health-only model showed a parallel pattern), and joint high chronic burden class membership overlapped substantially with both domain-specific high-burden classes (68.0\% and 63.8\%, respectively), supporting the interpretation that the joint classes reflect genuinely co-occurring, rather than artificially conflated, mental and physical health burden. A survey-weighted class-membership model, using the Wave V longitudinal weight, produced materially similar odds ratios (forced sexual experience, high chronic burden: OR = 3.00, 95\% CrI 2.29--3.96; moderate/worsening: OR = 1.34, 95\% CrI 1.04--1.74) to the primary unweighted model, and a parallel survey-weight descriptive sensitivity analysis found similar weighted and unweighted class proportions and exposure prevalence, indicating that complex survey design was unlikely to materially bias the primary findings. A complete-case sensitivity analysis using relationship-specific partner-forced-sex indicators from Waves III and IV, evaluated separately because of substantial missingness, supported the primary finding: partner forced sex was associated with higher posterior odds of high chronic burden membership (OR = 2.33, 95\% CrI 1.39--3.96). An exploratory equity analysis found a higher predicted probability of high chronic burden membership among Hispanic relative to non-Hispanic women (Supplementary Material), a finding requiring cautious interpretation given limited power for subgroup analysis.

\section*{Discussion}

Long-term mental and physical health following forced sexual experience was heterogeneous, following a low-burden/resilient, a moderate/worsening, or a high chronic burden course rather than a single average trajectory. This three-class structure was consistently recovered across three Bayesian mixture specifications with different priors and identification strategies -- including an unconstrained specification with no ordering imposed during sampling, relabeled only after estimation -- indicating that the structure was not an artifact of a single modeling approach or class-labeling procedure. The three-class solution also outperformed four- and five-class alternatives on convergence and class separation despite their nominally better BIC/AIC: the four-class model fragmented the moderate/worsening class into two statistically indistinguishable components (a signature of over-extraction that manifested directly as poor MCMC mixing), and the five-class model isolated a very small, weakly separated subgroup within the high chronic burden class. This pattern is consistent with recommended practice for finite mixture model selection that weighs convergence diagnostics and class separation alongside information criteria rather than relying on fit statistics alone (Muthén, 2010; Stephens, 2000).

This convergence-and-separation-based approach to class-count selection is, more generally, a methodological contribution that extends beyond this specific application. Information criteria such as BIC and AIC are widely used to select the number of classes in growth mixture and latent class analyses, but they can favor increasingly complex solutions even when additional classes reflect the fragmentation of genuine structure or the isolation of small, idiosyncratic subgroups rather than additional generalizable patterns, particularly in samples of the size available here. Directly comparing posterior convergence diagnostics and class separation across candidate values of $K$, as done here for $K=3,4,5$, provides an additional, complementary check that is straightforward to implement whenever a Bayesian mixture model is already being fit, and may be a useful addition to standard class-enumeration practice in future applications of Bayesian mixture modeling to trajectory data of this kind.

These findings extend prior posttraumatic trajectory research (Bonanno et al., 2011; Galatzer-Levy et al., 2018; Norris et al., 2009; Pietrzak et al., 2014) to a nationally representative, multi-decade cohort and to the joint course of mental and physical health specifically, rather than a single symptom domain. The magnitude of the association between forced sexual experience and high chronic burden membership -- nearly fourfold higher posterior odds relative to low-burden/resilient membership -- is consistent with, and quantitatively extends, prior evidence on the psychological and physical health consequences of sexual violence (Dworkin et al., 2017; Basile et al., 2021, 2024) and with the broader intimate partner violence literature documenting similarly elevated, co-occurring mental and physical health risk (Campbell, 2008; Bacchus et al., 2018; White et al., 2023). That physical and nonphysical forced sexual experience showed similarly elevated, independent associations with high chronic burden membership suggests that both experiences carry substantial long-term health consequences, consistent with uniform surveillance definitions that treat both as clinically meaningful sexual violence exposures (Basile et al., 2014).

From a clinical and public health standpoint, the identification of a high chronic burden class comprising nearly 30\% of the analytic sample, and disproportionately populated by women with forced sexual experience, argues against a one-size-fits-all model of post-assault care that assumes gradual improvement for all survivors. Instead, these findings support routine, longitudinal assessment of both mental and physical health among women with a history of forced sexual experience, with escalation of care intensity for those showing a chronic or worsening course rather than only those presenting with acute symptoms shortly after disclosure. Because forced sexual experience is common in this population-based sample (33.3\%) and its association with unfavorable trajectories is large, even modest improvements in healthcare access or social connectedness, if causally linked to trajectory membership, could plausibly translate into meaningful population-level reductions in chronic mental and physical health burden; this population-level framing, however, should be interpreted as motivation for further causal and intervention research rather than as an estimate of intervention efficacy.

Healthcare access and social connectedness were independently associated with lower posterior odds of unfavorable trajectory membership, complementing prior work on social reactions and support after sexual violence (Dworkin et al., 2019; Ullman, 2016; Howe et al., 2024; WHO, 2013) and pointing to two concrete, modifiable targets for trauma-informed care and public health intervention more broadly (SAMHSA, 2014). The absence of a clear moderation effect suggests these resources are associated with more favorable trajectories broadly, rather than specifically buffering the effect of forced sexual experience on trajectory membership -- a distinction relevant to intervention design, since it implies that improving healthcare access and social connectedness may benefit women's long-term health course independent of exposure history, rather than functioning only as an exposure-specific protective mechanism. The exploratory equity finding of higher predicted high chronic burden probability among Hispanic women warrants replication with adequately powered, prespecified equity models rather than immediate clinical extrapolation.

Several plausible mechanisms could link healthcare access and social connectedness to more favorable joint mental and physical health trajectories, although the present exploratory pathway analyses cannot distinguish among them. Better healthcare access may operate through earlier detection and treatment of both depressive symptoms and physical health conditions, reducing the likelihood that either domain deteriorates unchecked; it may also reflect broader socioeconomic resources that independently support recovery. Social connectedness may operate through practical and emotional support that buffers the psychological impact of forced sexual experience, through encouraging healthcare-seeking behavior, or through reducing the social isolation that has itself been linked to poorer physical health outcomes in the general population (Schnurr \& Green, 2004). Because mental and physical health were modeled jointly rather than separately, this study cannot determine whether healthcare access and social connectedness act primarily on one domain with secondary effects on the other, or independently on both; disentangling these pathways is an important direction for future research using designs with more closely spaced repeated measures of both health domains and candidate mediators.

This study has limitations. A central limitation concerns exposure timing: the joint trajectory indicators begin at Wave III, but the primary forced sexual experience measure is drawn from Wave IV and Wave V items, so some exposure may have occurred before Wave III. Consequently, part of what is modeled as a post-exposure trajectory could already reflect health status following an assault that occurred prior to the first trajectory measurement, and the Wave III indicators cannot be assumed to represent a true pre-exposure baseline for all women. This is a reverse-causation concern rather than a purely descriptive one, and it means that associations reported here should be interpreted as describing trajectory membership among women with a history of forced sexual experience by young adulthood, rather than as evidence that forced sexual experience causally precedes and produces the entire observed trajectory. Add Health's public-use forced sexual experience measure also combines physical and nonphysical experiences at Waves IV and V and cannot further resolve exposure timing within that window; relationship-specific partner-forced-sex items required a separate complete-case sensitivity analysis because of substantial missingness and coding ambiguity. Early adversity indicators of substantive interest were excluded from the adjusted models because of near-universal missingness in the public-use data, so this study cannot rule out that childhood adversity, rather than forced sexual experience specifically, accounts for part of the observed association with trajectory membership. The depressive symptom and self-rated physical health indicators available in Add Health, while suitable for population-based trajectory modeling, do not capture the full range of trauma-related sequelae documented in more specialized clinical instruments, and the two-item-per-domain structure limits the granularity with which mental and physical health burden can be characterized within each latent class. The pathway analyses linking healthcare access and social connectedness to trajectory membership were exploratory and do not support formal causal mediation claims, which would require additional assumptions about unmeasured exposure-mediator, exposure-outcome, and mediator-outcome confounding, together with clearer temporal ordering than the available waves permit (Imai et al., 2010; VanderWeele, 2015; Yuan \& MacKinnon, 2009). As with any observational analysis, residual confounding by unmeasured factors cannot be excluded, and the Bayesian mixture and class-membership models, however carefully specified, remain descriptive rather than causal accounts of trajectory membership. The complete-case restriction required to construct the joint trajectory indicators also excluded a substantial share of eligible women (see Data source and sample), and this attrition was not random with respect to race, ethnicity, and forced sexual experience, which may limit generalizability of the findings, particularly for Hispanic and Black or African American women, who were somewhat underrepresented in the final analytic sample relative to the eligible baseline population. Finally, although Add Health is nationally representative of the U.S.\ school-based adolescent population at baseline, attrition over five waves of follow-up and the public-use sample restrictions applied here may not fully preserve that representativeness in the present analytic sample. Future work should pursue formal counterfactual mediation analysis, prespecified and adequately powered equity moderation models, and validated, survivor-centered measures of healing and recovery designed specifically for gender-based violence survivors (Sinko et al., 2021, 2022), which may capture recovery dimensions not reflected in the depressive symptom and self-rated physical health indicators available in Add Health. Replication in independent cohorts, and eventually in restricted-use Add Health data with more detailed exposure timing, would further strengthen confidence in the trajectory structure identified here.

\section*{Conclusion}

Women's long-term mental and physical health following forced sexual experience is heterogeneous, not uniform, and this heterogeneity can be characterized by three distinct, robustly estimated joint trajectory classes: a low-burden/resilient class, a moderate/worsening burden class, and a high chronic burden class. This structure was recovered consistently across three Bayesian mixture specifications with different priors and identification strategies, and it was preferred over four- and five-class alternatives on convergence and class-separation grounds despite their nominally better fit statistics, providing a methodologically grounded answer to a question -- how many classes best describe recovery after forced sexual experience -- that is often settled by fit statistics alone. Forced sexual experience was consistently associated with membership in less favorable trajectories, most strongly the high chronic burden class, across every nested class-membership specification examined, and this association was neither attenuated nor amplified by the inclusion of healthcare access, social connectedness, or demographic covariates, indicating a robust rather than confounded relationship. Better healthcare access and greater social connectedness were, in turn, independently associated with more favorable trajectory membership, broadly rather than exposure-specifically. Taken together, these findings support a holistic recovery framework that integrates mental health, physical health, and modifiable post-assault resources, and they identify healthcare access and social connectedness as two concrete, actionable targets for trauma-informed clinical practice and public health intervention aimed at improving women's long-term health after sexual violence.

\vspace{4em}

\section*{Acknowledgement}
This research uses data from Add Health, a program project designed by J. Richard Udry, Peter S. Bearman, and Kathleen Mullan Harris, and funded by a grant P01-HD31921 from the Eunice Kennedy Shriver National Institute of Child Health and Human Development, with cooperative funding from 17 other agencies. Special acknowledgment is due Ronald R. Rindfuss and Barbara Entwisle for assistance in the original design. Persons interested in obtaining Data Files from Add Health should contact Add Health, The University of North Carolina at Chapel Hill, Carolina Population Center, Carolina Square, Suite 210, 123 W. Franklin Street, Chapel Hill, NC 27516 (addhealth\_contracts@unc.edu). No direct support was received from grant P01-HD31921 for this analysis. The views expressed in this manuscript are those of the authors and do not necessarily represent the views of the institutes or agencies that funded Add Health data collection or the University of North Carolina at Chapel Hill. Comments of the Editor and the anonymous reviewers are greatly acknowledged.

\section*{Funding}
This research received no specific grant from any funding agency, commercial or not-for-profit sectors.

\section*{Competing Interests}
The authors declare no known conflict of interest.

\section*{Ethical Standards}
The authors assert that all procedures contributing to this work comply with the ethical standards of the relevant national and institutional committees on human experimentation and with the Helsinki Declaration of 1975, as revised in 2008. Add Health data collection was approved by the Institutional Review Board of the University of North Carolina at Chapel Hill; the present analysis used de-identified public-use data.

\section*{Data Availability}
This study used public-use data from the National Longitudinal Study of Adolescent to Adult Health (Add Health). Add Health data are available subject to Add Health data access procedures and conditions of use and are available to researchers from the Carolina Population Center (\url{https://addhealth.cpc.unc.edu}). The analysis code (Python/PyMC) used to construct the analytic sample and estimate all models reported here is available on GitHub via \url{https://github.com/haolayinka/addhealth-bayesian-recovery-trajectories}.

\clearpage

\section*{References}
{\small
\begin{description}
\item[] Bacchus, L. J., Ranganathan, M., Watts, C., \& Devries, K. (2018). Recent intimate partner violence against women and health: A systematic review and meta-analysis of cohort studies. \textit{BMJ Open, 8}, e019995.
\item[] Basile, K. C., Smith, S. G., Breiding, M. J., Black, M. C., \& Mahendra, R. R. (2014). \textit{Sexual Violence Surveillance: Uniform Definitions and Recommended Data Elements}. Centers for Disease Control and Prevention.
\item[] Basile, K. C., Smith, S. G., Chen, J., \& Zwald, M. (2021). Chronic diseases, health conditions, and other impacts associated with rape victimization of U.S. women. \textit{Journal of Interpersonal Violence, 36}(23--24), NP12504--NP12520.
\item[] Basile, K. C., Chen, J., Friar, N. W., Smith, S. G., \& Leemis, R. W. (2024). Association of contact sexual violence victimization and health in the U.S. \textit{American Journal of Preventive Medicine, 66}(5), 860--869.
\item[] Bonanno, G. A. (2004). Loss, trauma, and human resilience. \textit{American Psychologist, 59}(1), 20--28.
\item[] Bonanno, G. A., Westphal, M., \& Mancini, A. D. (2011). Resilience to loss and potential trauma. \textit{Annual Review of Clinical Psychology, 7}, 511--535.
\item[] Campbell, J. C. (2008). The health consequences of intimate partner violence. \textit{The Lancet, 359}(9314), 1331--1336.
\item[] Campbell, R., Sprague, H. B., Cottrill, S., \& Sullivan, C. M. (2011). Longitudinal research with sexual assault survivors. \textit{Journal of Interpersonal Violence, 26}(3), 433--461.
\item[] Dworkin, E. R., Menon, S. V., Bystrynski, J., \& Allen, N. E. (2017). Sexual assault victimization and psychopathology: A review and meta-analysis. \textit{Clinical Psychology Review, 56}, 65--81.
\item[] Dworkin, E. R., Brill, C. D., \& Ullman, S. E. (2019). Social reactions to disclosure of interpersonal violence and psychopathology. \textit{Clinical Psychology Review, 72}, 101750.
\item[] Galatzer-Levy, I. R., Huang, S. H., \& Bonanno, G. A. (2018). Trajectories of resilience and dysfunction following potential trauma. \textit{Clinical Psychology: Science and Practice, 25}(2), e12241.
\item[] Garcia-Moreno, C., Jansen, H. A. F. M., Ellsberg, M., Heise, L., \& Watts, C. (2006). Prevalence of intimate partner violence. \textit{The Lancet, 368}(9543), 1260--1269.
\item[] Goodman-Williams, R., Clark, S. L., Campbell, R., \& Ullman, S. E. (2024). Longitudinal patterns of PTSD symptoms among sexual assault survivors. \textit{Psychological Trauma, 16}(4), 586--595.
\item[] Halfon, N., Larson, K., Lu, M., Tullis, E., \& Russ, S. (2014). Lifecourse health development: Past, present and future. \textit{Maternal and Child Health Journal, 18}(2), 344--365.
\item[] Griffiths, A.M., Hummer, R.A., Harris, K.M., Dean, S.C., Jackson, C. (2026). Understanding Sample Sizes, Sample Composition, and Cross-Sectional Sampling Weights: Add Health Survey Waves I-VI. Chapel Hill, NC: Carolina Population Center, University of North Carolina at Chapel Hill.
\item[] Howe, E. S., \& Dworkin, E. R. (2024). The day-to-day relationship between posttraumatic stress symptoms and social support after sexual assault. \textit{European Journal of Psychotraumatology, 15}(1), 2311478.
\item[] Imai, K., Keele, L., \& Tingley, D. (2010). A general approach to causal mediation analysis. \textit{Psychological Methods, 15}(4), 309--334.
\item[] Muthen, B. (2002). Beyond SEM: General latent variable modeling. \textit{Behaviormetrika, 29}, 81--117.
\item[] Muthen, B., \& Asparouhov, T. (2009). Growth mixture modeling: Analysis with non-Gaussian random effects. In \textit{Longitudinal Data Analysis}. Chapman \& Hall/CRC.
\item[] Muthen, B. (2010). Bayesian analysis in Mplus: A brief introduction. \textit{Mplus Technical Report}.
\item[] Norris, F. H., Tracy, M., \& Galea, S. (2009). Looking for resilience. \textit{Social Science \& Medicine, 68}(12), 2190--2198.
\item[] Olalude, G. A., Olayinka, H. A., Esiegbe, E. O., \& Adebiyi, A. A. (2025). Justification of intimate partner violence among women in Nigeria: how does it differ by socioeconomic status?. Journal of interpersonal violence, 40(13-14), 3085-3107.
\item[] Pearlin, L. I., Menaghan, E. G., Lieberman, M. A., \& Mullan, J. T. (1981). The stress process. \textit{Journal of Health and Social Behavior, 22}(4), 337--356.
\item[] Pietrzak, R. H., Feder, A., Singh, R., et al. (2014). Trajectories of PTSD risk and resilience. \textit{Psychological Medicine, 44}(1), 205--219.
\item[] Schnurr, P. P., \& Green, B. L. (2004). Understanding relationships among trauma, PTSD, and health outcomes. \textit{Advances in Mind-Body Medicine, 20}(1), 18--29.
\item[] Sinko, L., Schaitkin, C., \& Saint Arnault, D. (2021). The Healing after Gender-Based Violence Scale. \textit{Global Qualitative Nursing Research, 8}, 2333393621996679.
\item[] Sinko, L., Ozaslan, Z. Z., Schaitkin, C., \& Saint Arnault, D. (2022). Psychometric evaluation of the Healing after Gender-Based Violence Scale. \textit{Journal of Family Violence, 37}, 1161--1179.
\item[] Short, N. A., Sullivan, J., Soward, A., et al. (2019). Protocol for a large-scale longitudinal cohort study of recovery after sexual assault. \textit{BMJ Open, 9}, e031087.
\item[] Short, N. A., Morabito, D. M., \& Gilmore, A. K. (2020). Secondary prevention after recent sexual assault. \textit{Depression and Anxiety, 37}(10), 1047--1059.
\item[] Smith, S. G., Zhang, X., Basile, K. C., et al. (2018). \textit{The National Intimate Partner and Sexual Violence Survey: 2015 Data Brief}. Centers for Disease Control and Prevention.
\item[] Steenkamp, M. M., Dickstein, B. D., Salters-Pedneault, K., Hofmann, S. G., \& Litz, B. T. (2012). Trajectories of PTSD symptoms following sexual assault. \textit{Journal of Traumatic Stress, 25}(4), 469--474.
\item[] Stephens, M. (2000). Dealing with label switching in mixture models. \textit{Journal of the Royal Statistical Society: Series B, 62}(4), 795--809.
\item[] Substance Abuse and Mental Health Services Administration. (2014). \textit{SAMHSA's Concept of Trauma and Guidance for a Trauma-Informed Approach}.
\item[] Ullman, S. E., \& Peter-Hagene, L. C. (2016). Longitudinal relationships of social reactions, PTSD, and revictimization. \textit{Journal of Interpersonal Violence, 31}(6), 1074--1094.
\item[] VanderWeele, T. J. (2015). \textit{Explanation in Causal Inference}. Oxford University Press.
\item[] White, S. J., Sin, J., Sweeney, A., et al. (2023). Global prevalence and mental health outcomes of intimate partner violence. \textit{Trauma, Violence, \& Abuse, 25}(1), 494--511.
\item[] World Health Organization. (2013). \textit{Responding to Intimate Partner Violence and Sexual Violence Against Women}. WHO.
\item[] World Health Organization. (2021). \textit{Violence Against Women Prevalence Estimates, 2018}. WHO.
\item[] Yuan, Y., \& MacKinnon, D. P. (2009). Bayesian mediation analysis. \textit{Psychological Methods, 14}(4), 301--322.
\end{description}
}

\clearpage

\begin{table}[H]
\centering
\setlength{\tabcolsep}{4pt}
\begin{threeparttable}
\caption{Sample characteristics of the final analytic sample, overall and by forced sexual experience status.}
\footnotesize
\begin{tabular}{lccc}
\toprule
Characteristic & Overall (N=1{,}848) & No FSE (N=1{,}232) & FSE (N=616) \\
\midrule
Age, Wave III, mean (SD) & 21.64 (1.78) & 21.68 (1.78) & 21.56 (1.78) \\
Healthcare access, Wave III, mean (SD) & 0.81 (0.39) & 0.83 (0.35) & 0.76 (0.44) \\
Healthcare access, Wave IV, mean (SD) & 0.98 (0.10) & 0.98 (0.11) & 0.98 (0.09) \\
Close friends, Wave IV, mean (SD) & 3.14 (0.94) & 3.15 (0.94) & 3.12 (0.94) \\
Depressive symptom burden, Wave III, mean (SD) & 0.48 (0.57) & 0.42 (0.51) & 0.60 (0.64) \\
Depressive symptom burden, Wave V, mean (SD) & 1.10 (0.61) & 1.01 (0.55) & 1.27 (0.69) \\
Physical health burden, Wave III, mean (SD) & 2.06 (0.89) & 1.96 (0.87) & 2.25 (0.90) \\
Physical health burden, Wave V, mean (SD) & 2.45 (0.98) & 2.37 (0.95) & 2.62 (1.00) \\
Physical forced sexual experience, n (\%) & 366 (19.8) & 0 (0.0) & 366 (59.4) \\
Nonphysical forced sexual experience, n (\%) & 550 (29.8) & 0 (0.0) & 550 (89.3) \\
Hispanic ethnicity, n (\%) & 169 (9.1) & 118 (9.6) & 51 (8.3) \\
White, n (\%) & 1318 (71.8) & 855 (69.9) & 463 (75.8) \\
Black or African American, n (\%) & 450 (24.4) & 315 (25.6) & 135 (21.9) \\
American Indian or Native American, n (\%) & 78 (4.2) & 49 (4.0) & 29 (4.7) \\
Asian, n (\%) & 74 (4.0) & 54 (4.4) & 20 (3.2) \\
\bottomrule
\end{tabular}
\begin{tablenotes}[flushleft]
\footnotesize
\item \textbf{Note.} FSE = forced sexual experience. Race categories are non-mutually-exclusive binary indicators (participants may endorse more than one). Depressive symptom burden and physical health burden are shown for Waves III and V only here; full Wave IV values and missingness by variable are reported in Supplementary Material. Race and ethnicity are self-reported.
\end{tablenotes}
\end{threeparttable}
\end{table}

\begin{table}[H]
\centering
\begin{threeparttable}
\caption{Fully adjusted Bayesian class-membership model: posterior odds ratios for trajectory class membership relative to the low-burden/resilient class.}
\small
\begin{tabular}{lcccc}
\toprule
& \multicolumn{2}{c}{Moderate/worsening} & \multicolumn{2}{c}{High chronic} \\
Predictor & OR & 95\% CrI & OR & 95\% CrI \\
\midrule
Any forced sexual experience & 1.63 & (1.26, 2.12) & 3.78 & (2.84, 5.04) \\
Age, Wave III & 0.98 & (0.88, 1.10) & 0.94 & (0.83, 1.08) \\
Hispanic ethnicity & 1.05 & (0.67, 1.68) & 1.72 & (1.07, 2.82) \\
White & 1.07 & (0.57, 2.02) & 1.06 & (0.48, 2.37) \\
Black or African American & 1.61 & (0.86, 3.13) & 2.38 & (1.08, 5.47) \\
American Indian or Native American & 1.53 & (0.81, 3.03) & 1.25 & (0.59, 2.67) \\
Asian & 1.40 & (0.69, 2.99) & 0.97 & (0.39, 2.44) \\
Healthcare access, Wave III & 0.85 & (0.75, 0.96) & 0.65 & (0.56, 0.76) \\
Healthcare access, Wave IV & 1.02 & (0.91, 1.14) & 1.09 & (0.95, 1.27) \\
Close friends, Wave IV & 0.88 & (0.78, 0.98) & 0.70 & (0.60, 0.80) \\
\bottomrule
\end{tabular}
\begin{tablenotes}[flushleft]
\footnotesize
\item \textbf{Note.} OR = posterior median odds ratio. CrI = credible interval. Race categories are coded as separate non-mutually-exclusive binary indicators, each contrasted against not-endorsing that category, consistent with the underlying Add Health item coding; ``White,'' ``Black or African American,'' ``American Indian or Native American,'' and ``Asian'' correspond to the race variables reported in the original data. $N=1{,}805$.
\end{tablenotes}
\end{threeparttable}
\end{table}

\begin{figure}[H]
\centering
\includegraphics[width=0.8\linewidth]{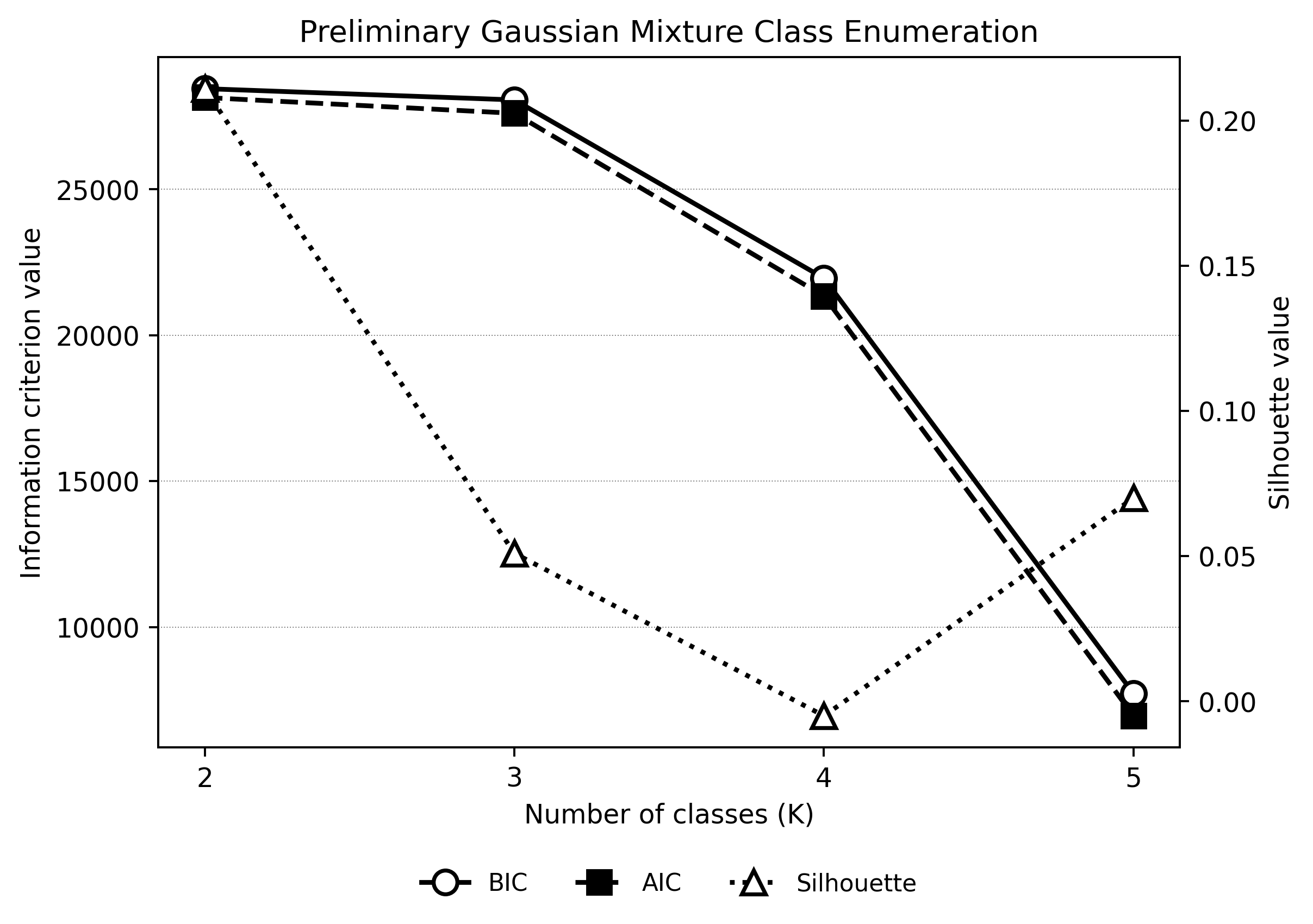}
\caption{Preliminary Gaussian mixture class enumeration: BIC, AIC, and silhouette values by number of classes.}
\end{figure}

\begin{figure}[H]
\centering
\includegraphics[width=0.95\linewidth]{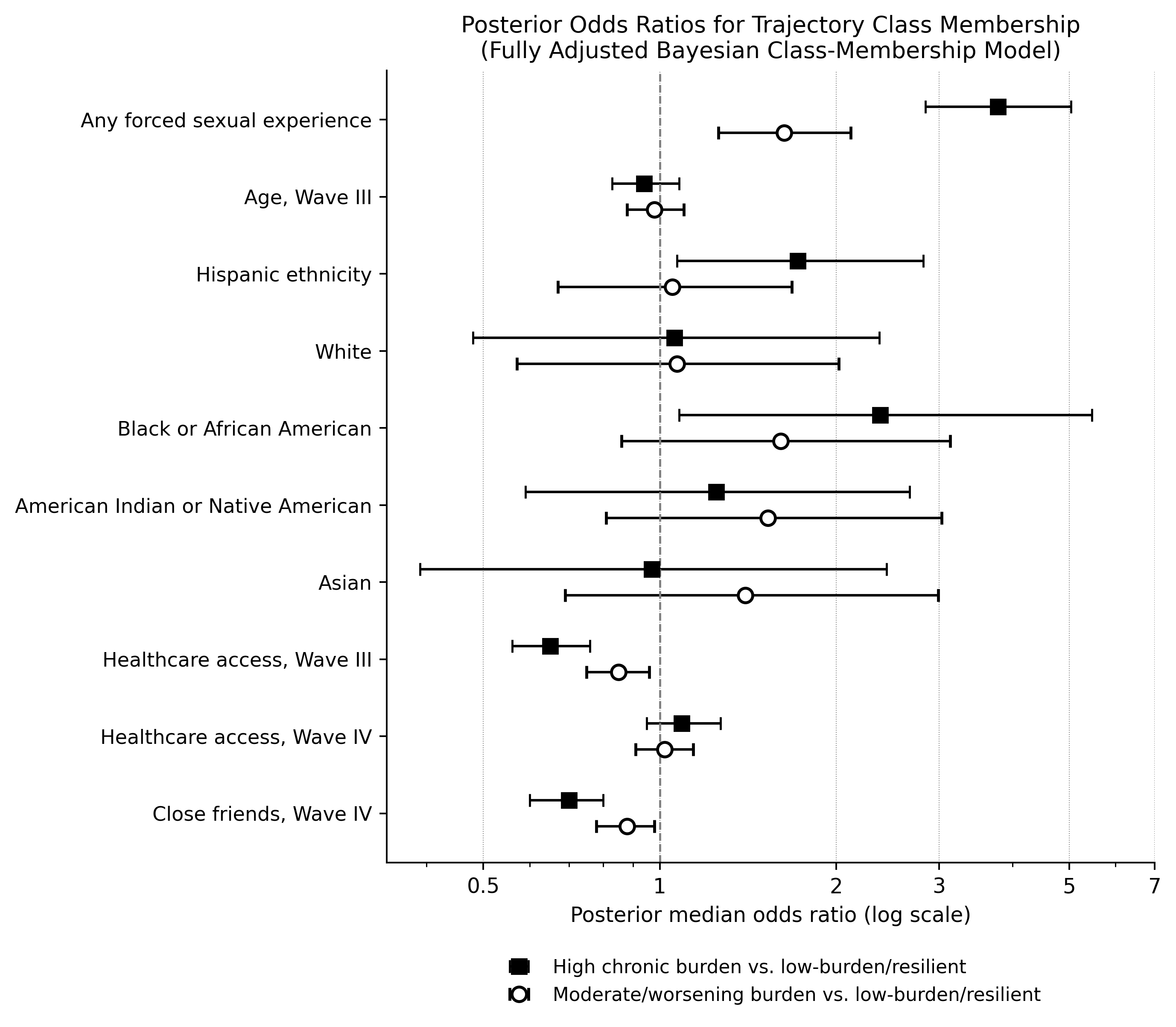}
\caption{Posterior odds ratios (95\% credible intervals) for trajectory class membership, fully adjusted Bayesian class-membership model.}
\end{figure}

\begin{figure}[H]
\centering
\includegraphics[width=0.85\linewidth]{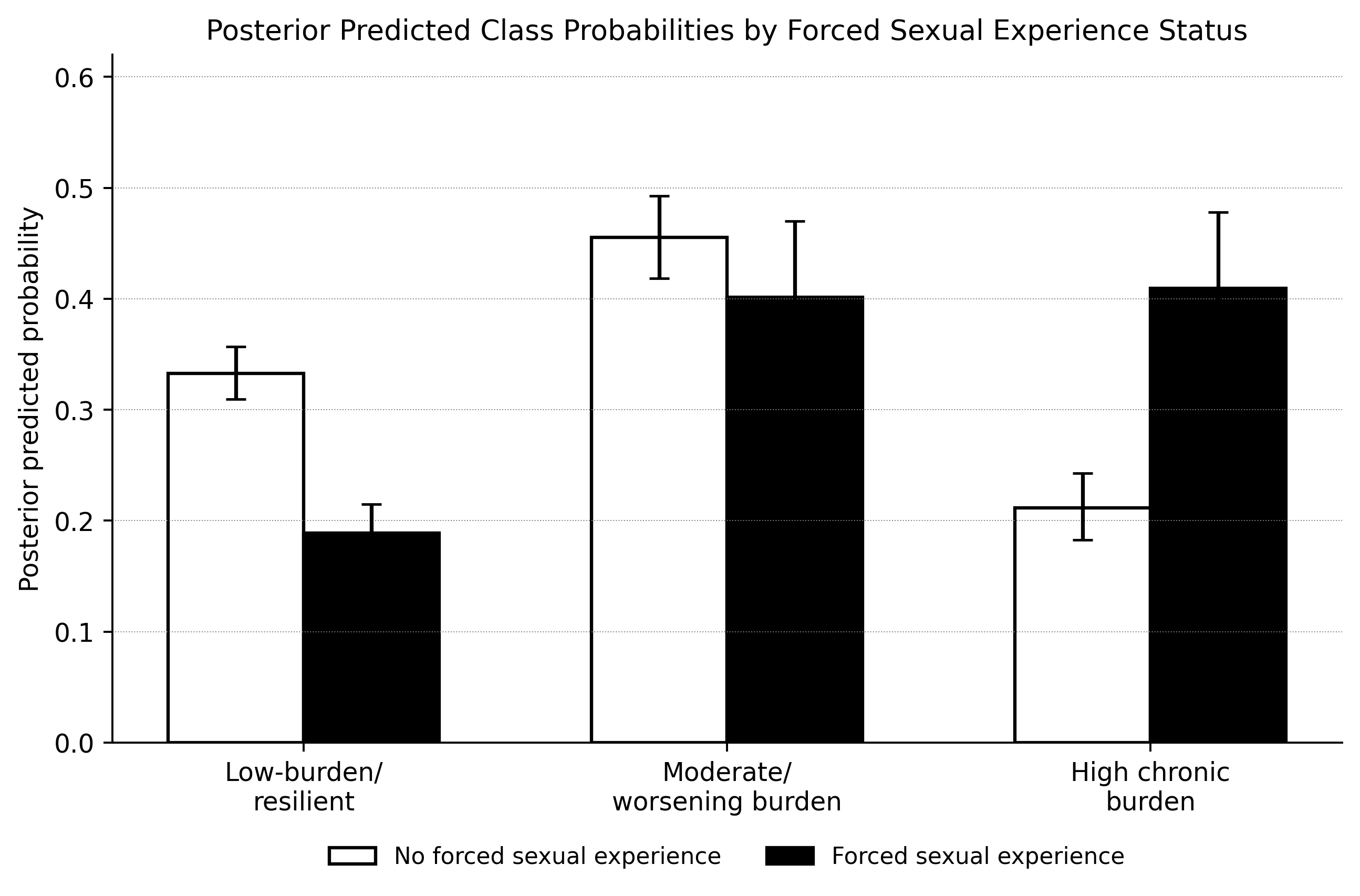}
\caption{Posterior predicted class probabilities by forced sexual experience status, fully adjusted Bayesian class-membership model, evaluated at the average covariate profile. Error bars represent 95\% credible intervals.}
\end{figure}

\end{document}

\maketitle

\renewcommand{\thesection}{S\arabic{section}}
\renewcommand{\thetable}{S\arabic{table}}
\renewcommand{\thefigure}{S\arabic{figure}}

\noindent This Supplementary Material accompanies the main manuscript. It reports full model specifications, posterior summaries, and sensitivity analyses that were referenced but not reproduced in full in the main text because of journal length and display-item limits.

\section{Sample Derivation and Comparison of Included Versus Excluded Women}

The analytic sample was derived from 2,629 female Add Health respondents with valid Wave III data. Table~S1 shows the step-by-step attrition to the final trajectory-complete analytic sample (N=1,848) and the further attrition to the fully adjusted class-membership sample (N=1,805).

\begin{table}[h]
\centering
\begin{threeparttable}
\caption{Sample derivation.}
\small
\begin{tabular}{lc}
\toprule
Sample restriction step & n remaining (n excluded at this step) \\
\midrule
Female respondents with valid Wave III data & 2{,}629 \\
Requiring Wave III depressive symptom data & 2{,}629 (0) \\
Requiring Wave IV depressive symptom data & 2{,}333 (296) \\
Requiring Wave V depressive symptom data & 1{,}850 (483) \\
Requiring Wave III physical health data & 1{,}850 (0) \\
Requiring Wave IV physical health data & 1{,}850 (0) \\
Requiring Wave V physical health data & 1{,}848 (2) \\
\textbf{Final trajectory-complete analytic sample} & \textbf{1{,}848} \\
Requiring Wave IV healthcare access & 1{,}839 (9) \\
Requiring Wave IV closeness to friends & 1{,}818 (21) \\
Requiring Hispanic ethnicity and race data & 1{,}805 (13) \\
\textbf{Final fully adjusted class-membership sample} & \textbf{1{,}805} \\
\bottomrule
\end{tabular}
\begin{tablenotes}[flushleft]
\footnotesize
\item \textbf{Note.} Trajectory indicators are depressive symptom burden and self-rated physical health burden. Adjusted-model covariates additionally required are Wave IV healthcare access, Wave IV closeness to friends, Hispanic ethnicity, and race (White, Black or African American, American Indian or Native American, Asian).
\end{tablenotes}
\end{threeparttable}
\end{table}

Table~S2 compares baseline (Wave III) characteristics of the 1,848 women retained in the trajectory-complete analytic sample against the 781 women excluded for incomplete trajectory data, using independent-samples $t$-tests for continuous variables and chi-square tests for categorical variables.

\begin{table}[h]
\centering
\begin{threeparttable}
\caption{Comparison of included versus excluded women on baseline (Wave III) characteristics.}
\small
\begin{tabular}{lccc}
\toprule
Characteristic & Included (n=1{,}848) & Excluded (n=781) & $p$ \\
\midrule
Age, Wave III, mean (SD) & 21.64 (1.78) & 21.93 (1.84) & $<.001$ \\
Any forced sexual experience, \% & 33.3 & 23.1 & $<.001$ \\
Hispanic ethnicity, \% & 9.1 & 13.9 & $<.001$ \\
White, \% & 71.8 & 62.2 & $<.001$ \\
Black or African American, \% & 24.4 & 30.3 & .002 \\
Healthcare access, Wave III, mean (SD) & 0.81 (0.39) & 0.79 (0.39) & .33 \\
\bottomrule
\end{tabular}
\begin{tablenotes}[flushleft]
\footnotesize
\item \textbf{Note.} Excluded women were missing data on at least one of the six joint trajectory indicators (depressive symptom burden or self-rated physical health burden at Waves III, IV, or V). $p$-values are from independent-samples $t$-tests (continuous variables) or chi-square tests (categorical variables). Forced sexual experience, Hispanic ethnicity, White, and Black or African American reflect Wave III/IV/V items with their own item-level missingness among excluded women; percentages are calculated among those with non-missing data on each variable.
\end{tablenotes}
\end{threeparttable}
\end{table}

\begin{sloppypar}
Gaussian mixture models with two through five classes were fit to the six standardized joint trajectory indicators (depressive symptom burden and self-rated physical health burden at Waves III, IV, and V) using \texttt{scikit-learn}'s \texttt{GaussianMixture} (full covariance, \texttt{random\_state=123}, \texttt{n\_init=30}). Table~S3 reports BIC, AIC, silhouette values, and class sizes for each candidate solution. This stage was used only to inform priors for the subsequent Bayesian mixture model, not as the final trajectory model.
\end{sloppypar}

\begin{table}[h]
\centering
\begin{threeparttable}
\caption{Preliminary Gaussian mixture class enumeration.}
\begin{tabular}{ccccl}
\toprule
$K$ & BIC & AIC & Silhouette & Class size proportions \\
\midrule
2 & 28448.45 & 28144.75 & 0.211 & 0.410, 0.590 \\
3 & 28062.27 & 27603.95 & 0.051 & 0.365, 0.310, 0.325 \\
4 & 21953.30 & 21340.38 & -0.005 & 0.160, 0.326, 0.172, 0.343 \\
5 & 7722.41 & 6954.88 & 0.070 & 0.050, 0.412, 0.298, 0.006, 0.233 \\
\bottomrule
\end{tabular}
\begin{tablenotes}[flushleft]
\footnotesize
\item \textbf{Note.} BIC and AIC continued to decrease through five classes; however, the five-class solution produced a very small class representing only 0.6\% of the sample. A three-class solution was carried forward to the identified Bayesian mixture model on substantive and diagnostic grounds (Section~S3).
\end{tablenotes}
\end{threeparttable}
\end{table}

\section{Full Bayesian Mixture Model Specifications}

Three Bayesian mixture specifications were estimated in PyMC 5.28.5 (four chains, 2,000 tuning and 4,000 post-tuning draws, target\_accept = 0.97).

\textbf{Identified model.} Let $\hat p_k$, $\hat\mu_{kj}$, $\hat\sigma_{kj}$ denote the preliminary Gaussian-mixture class proportions, means, and standard deviations (Section~S1), ordered by severity.
\begin{align*}
\boldsymbol{\pi} &\sim \text{Dirichlet}(\alpha_k), \quad \alpha_k = \max(40\hat p_k, 1.5),\\
\mu_{kj} &\sim N(\hat\mu_{kj}, 0.30^2),\\
\sigma_{kj} &\sim \text{TruncatedNormal}\left(\max(\hat\sigma_{kj},0.45),\,0.20^2;\,0.30,\,3.00\right),
\end{align*}
with an ordered class-severity constraint, $S_1<S_2<S_3$ where $S_k = \tfrac{1}{6}\sum_j \mu_{kj}$, imposed as a hard log-posterior penalty ($-10^{9}$ when violated).

\textbf{Weakly identified model.} Priors used only the rank order of severity, not data-derived class-specific values:
\begin{align*}
\boldsymbol{\pi} &\sim \text{Dirichlet}(1,1,1), \qquad \mu_{kj} \sim N(c_k, 1.25^2),\quad (c_1,c_2,c_3)=(-0.45,0.00,0.45),\\
\sigma_{kj} &\sim \text{TruncatedNormal}(1.00, 0.75^2; 0.30, 4.00),
\end{align*}
with the same ordered-severity constraint as the identified model.

\textbf{Unconstrained model.} Fully symmetric priors, no ordering constraint during sampling:
\begin{align*}
\boldsymbol{\pi} &\sim \text{Dirichlet}(1,1,1), \qquad \mu_{kj} \sim N(0, 2.00^2), \qquad \sigma_{kj} \sim \text{TruncatedNormal}(1.00, 0.75^2; 0.30, 4.00).
\end{align*}
Posterior draws were relabeled post hoc by class severity to resolve label-switching.

Table~S4 reports full posterior summaries for the identified model (the primary specification).

\begin{table}[h]
\centering
\setlength{\tabcolsep}{14pt}
\begin{threeparttable}
\caption{Identified Bayesian mixture model: full posterior summary.}
\footnotesize
\begin{tabular}{lccc}
\toprule
Parameter & Mean & SD/95\% CrI & $\hat{R}$ \\
\midrule
Class weight, low-burden/resilient & 0.272 & 0.014 (0.245, 0.301) & 1.00 \\
Class weight, moderate/worsening burden & 0.432 & 0.020 (0.394, 0.472) & 1.00 \\
Class weight, high chronic burden & 0.297 & 0.018 (0.261, 0.330) & 1.00 \\
Severity, low-burden/resilient & -0.539 & 0.018 (-0.574, -0.502) & 1.00 \\
Severity, moderate/worsening burden & -0.153 & 0.023 (-0.198, -0.109) & 1.00 \\
Severity, high chronic burden & 0.717 & 0.030 (0.656, 0.773) & 1.00 \\
\midrule
\multicolumn{4}{l}{\textit{Class-specific means, mental health burden (Waves III, IV, V)}} \\
Low-burden/resilient & \multicolumn{3}{l}{-0.655, -0.727, -0.707 (all $\hat R=1.00$)} \\
Moderate/worsening burden & \multicolumn{3}{l}{-0.070, -0.065, -0.093 (all $\hat R=1.00$)} \\
High chronic burden & \multicolumn{3}{l}{0.705, 0.763, 0.787 (all $\hat R=1.00$)} \\
\multicolumn{4}{l}{\textit{Class-specific means, physical health burden (Waves III, IV, V)}} \\
Low-burden/resilient & \multicolumn{3}{l}{-0.307, -0.395, -0.443 (all $\hat R=1.00$)} \\
Moderate/worsening burden & \multicolumn{3}{l}{-0.242, -0.232, -0.216 (all $\hat R=1.00$)} \\
High chronic burden & \multicolumn{3}{l}{0.631, 0.698, 0.719 (all $\hat R=1.00$)} \\
\bottomrule
\end{tabular}
\begin{tablenotes}[flushleft]
\footnotesize
\item \textbf{Note.} CrI = credible interval. Full class-specific standard deviations ($\sigma_{kj}$) and MCSE are omitted here for space and are available from the authors on request. Minimum effective sample size across all reported parameters was 2,068.
\end{tablenotes}
\end{threeparttable}
\end{table}

\section{Sensitivity to the Number of Latent Classes}

The identified Bayesian mixture model was re-estimated with four and five classes, using an identical specification to the primary three-class model. Table~S5 compares convergence and separation across $K=3,4,5$.

\begin{table}[h]
\centering
\setlength{\tabcolsep}{10pt}
\renewcommand{\arraystretch}{1.15}
\begin{threeparttable}
\caption{Sensitivity of the identified Bayesian mixture model to the number of latent classes.}
\small
\begin{tabular}{ccccc}
\toprule
Classes & Smallest class proportion & Max. $\hat{R}$ & Min. ESS & Min. severity gap \\
\midrule
3 (primary) & 0.272 & 1.00 & 2,068 & 0.386 \\
4 & 0.176 & 1.07 & 42 & 0.016 \\
5 & 0.027 & 1.00 & 1,176 & 0.066 \\
\bottomrule
\end{tabular}
\begin{tablenotes}[flushleft]
\footnotesize
\item \textbf{Note.} ESS = minimum bulk effective sample size across reported parameters from a 2000/2000/4 chains run. Minimum severity gap is the smallest difference between adjacent ordered class-severity posterior means.
\end{tablenotes}
\end{threeparttable}
\end{table}

\begin{table}[h]
\centering
\setlength{\tabcolsep}{12pt}
\renewcommand{\arraystretch}{1.15}
\begin{threeparttable}
\caption{Four-class identified Bayesian mixture model: posterior class weights and severity.}
\small
\begin{tabular}{lcccc}
\toprule
Class (ordered by severity) & PM & PSD & 95\% CrI & $\hat{R}$ \\
\midrule
Class weight 1 & 0.176 & 0.013 & (0.151, 0.199) & 1.07 \\
Class weight 2 & 0.242 & 0.023 & (0.198, 0.280) & 1.05 \\
Class weight 3 & 0.326 & 0.026 & (0.279, 0.377) & 1.06 \\
Class weight 4 & 0.256 & 0.015 & (0.229, 0.285) & 1.01 \\
Severity 1 & -0.737 & 0.020 & (-0.776, -0.702) & 1.06 \\
Severity 2 & -0.137 & 0.018 & (-0.172, -0.105) & 1.03 \\
Severity 3 & -0.121 & 0.017 & (-0.152, -0.089) & 1.03 \\
Severity 4 & 0.792 & 0.032 & (0.728, 0.849) & 1.03 \\
\bottomrule
\end{tabular}
\begin{tablenotes}[flushleft]
\footnotesize
\item \textbf{Note.} PM = posterior mean; PSD = posterior standard deviation. Classes 2 and 3 have severity estimates only 0.016 apart, within one posterior SD, consistent with the elevated $\hat{R}$ and low ESS (minimum 42) observed for this specification.
\end{tablenotes}
\end{threeparttable}
\end{table}

\begin{table}[h]
\centering
\setlength{\tabcolsep}{12pt}
\renewcommand{\arraystretch}{1.15}
\begin{threeparttable}
\caption{Five-class identified Bayesian mixture model: posterior class weights and severity.}
\small
\begin{tabular}{lcccc}
\toprule
Class (ordered by severity) & PM & PSD & 95\% CrI & $\hat{R}$ \\
\midrule
Class weight 1 & 0.255 & 0.011 & (0.235, 0.275) & 1.00 \\
Class weight 2 & 0.311 & 0.012 & (0.288, 0.334) & 1.00 \\
Class weight 3 & 0.142 & 0.009 & (0.124, 0.159) & 1.00 \\
Class weight 4 & 0.027 & 0.005 & (0.019, 0.036) & 1.00 \\
Class weight 5 & 0.265 & 0.014 & (0.239, 0.292) & 1.00 \\
Severity 1 & -0.570 & 0.014 & (-0.596, -0.543) & 1.00 \\
Severity 2 & -0.245 & 0.015 & (-0.272, -0.218) & 1.00 \\
Severity 3 & 0.111 & 0.020 & (0.076, 0.150) & 1.00 \\
Severity 4 & 0.656 & 0.046 & (0.564, 0.734) & 1.00 \\
Severity 5 & 0.722 & 0.026 & (0.672, 0.771) & 1.00 \\
\bottomrule
\end{tabular}
\begin{tablenotes}[flushleft]
\footnotesize
\item \textbf{Note.} Class 4 represents approximately 2.7\% of the sample ($\approx$50 women) with a severity estimate only 0.066 from Class 5, consistent with an idiosyncratic small subgroup within the high chronic burden class rather than a distinct trajectory.
\end{tablenotes}
\end{threeparttable}
\end{table}

\section{Classification Quality and Class Profile Checks}

The mean maximum posterior class probability was 0.865 (median 0.931); 72.2\% of respondents had maximum posterior class probability $\geq$0.80, and 56.8\% had probability $\geq$0.90. Classification certainty was highest for the high chronic burden class (mean 0.891) relative to the moderate/worsening (0.854) and low-burden/resilient (0.856) classes. Observed-versus-model-implied class profile discrepancies were small throughout: mean absolute differences of 0.008 (low-burden/resilient), 0.015 (moderate/worsening), and 0.052 (high chronic burden); maximum absolute differences of 0.019, 0.028, and 0.069, respectively.

\section{Full Class-Membership Model Results (Models 1--4)}

Four nested Bayesian class-membership models were estimated: Model 1 (forced sexual experience only), Model 2 (+ age), Model 3 (+ healthcare access and close friends), and Model 4 (+ Hispanic ethnicity and race indicators; the fully adjusted model reported in the main text). As presented in Table~S8, all four models showed a consistent, dose-like pattern of increasing posterior odds of less-favorable class membership with forced sexual experience.

\begin{ThreePartTable}
\begin{TableNotes}[flushleft]
\footnotesize
\item \textit{Note.} Class 1 = moderate/worsening burden trajectory. Class 2 = high chronic burden trajectory. Class 3 = low-burden/resilient reference trajectory. Models use final class assignments from the identified Bayesian mixture model. Model 1 adjusted for forced sexual experience and age. Model 2 added Wave III and Wave IV healthcare access. Model 3 added close friends at Wave IV. Model 4 additionally adjusted for Hispanic ethnicity and race indicators. Forced sexual experience and demographic indicators were entered as binary predictors. Age, healthcare access, and close friends were standardized; their odds ratios represent a one-standard-deviation increase. OR = posterior median odds ratio. CrI = credible interval.
\end{TableNotes}

\begingroup
\small
\setlength{\tabcolsep}{16pt}
\renewcommand{\arraystretch}{1.25}

\begin{longtable}{p{3.0cm}p{1.4cm}p{4.8cm}p{1.0cm}p{1.8cm}}
\caption{Final Bayesian class-membership models using identified Bayesian mixture class assignments.}
\label{tab:final_bayes_class_membership}\\

\toprule
Contrast & Model & Predictor & OR & 95\% CrI \\
\midrule
\endfirsthead

\multicolumn{5}{c}{\textit{Table~\ref{tab:final_bayes_class_membership} continued}}\\
\toprule
Contrast & Model & Predictor & OR & 95\% CrI \\
\midrule
\endhead

\midrule
\multicolumn{5}{r}{\textit{Continued on next page}}\\
\midrule
\endfoot

\bottomrule
\insertTableNotes
\endlastfoot

Class 1 vs Class 3 & Model 1 & Forced sexual experience & 1.55 & (1.20, 2.01) \\
Class 1 vs Class 3 & Model 1 & Age, Wave III & 0.99 & (0.88, 1.11) \\

\midrule
Class 1 vs Class 3 & Model 2 & Forced sexual experience & 1.55 & (1.20, 2.01) \\
Class 1 vs Class 3 & Model 2 & Age, Wave III & 0.98 & (0.87, 1.09) \\
Class 1 vs Class 3 & Model 2 & Healthcare access, Wave III & 0.84 & (0.75, 0.95) \\
Class 1 vs Class 3 & Model 2 & Healthcare access, Wave IV & 1.04 & (0.92, 1.16) \\

\midrule
Class 1 vs Class 3 & Model 3 & Forced sexual experience & 1.55 & (1.20, 2.02) \\
Class 1 vs Class 3 & Model 3 & Age, Wave III & 0.97 & (0.87, 1.09) \\
Class 1 vs Class 3 & Model 3 & Healthcare access, Wave III & 0.86 & (0.76, 0.96) \\
Class 1 vs Class 3 & Model 3 & Healthcare access, Wave IV & 1.04 & (0.93, 1.17) \\
Class 1 vs Class 3 & Model 3 & Close friends, Wave IV & 0.86 & (0.77, 0.96) \\

\midrule
Class 1 vs Class 3 & Model 4 & Forced sexual experience & 1.63 & (1.26, 2.12) \\
Class 1 vs Class 3 & Model 4 & Age, Wave III & 0.98 & (0.88, 1.10) \\
Class 1 vs Class 3 & Model 4 & Hispanic ethnicity & 1.05 & (0.67, 1.68) \\
Class 1 vs Class 3 & Model 4 & White & 1.07 & (0.57, 2.02) \\
Class 1 vs Class 3 & Model 4 & Black or African American & 1.61 & (0.86, 3.13) \\
Class 1 vs Class 3 & Model 4 & American Indian or Native American & 1.53 & (0.81, 3.03) \\
Class 1 vs Class 3 & Model 4 & Asian & 1.40 & (0.69, 2.99) \\
Class 1 vs Class 3 & Model 4 & Healthcare access, Wave III & 0.85 & (0.75, 0.96) \\
Class 1 vs Class 3 & Model 4 & Healthcare access, Wave IV & 1.02 & (0.91, 1.14) \\
Class 1 vs Class 3 & Model 4 & Close friends, Wave IV & 0.88 & (0.78, 0.98) \\

\midrule
Class 2 vs Class 3 & Model 1 & Forced sexual experience & 3.34 & (2.58, 4.41) \\
Class 2 vs Class 3 & Model 1 & Age, Wave III & 0.95 & (0.84, 1.08) \\

\midrule
Class 2 vs Class 3 & Model 2 & Forced sexual experience & 3.37 & (2.56, 4.42) \\
Class 2 vs Class 3 & Model 2 & Age, Wave III & 0.94 & (0.83, 1.07) \\
Class 2 vs Class 3 & Model 2 & Healthcare access, Wave III & 0.65 & (0.56, 0.75) \\
Class 2 vs Class 3 & Model 2 & Healthcare access, Wave IV & 1.11 & (0.97, 1.29) \\

\midrule
Class 2 vs Class 3 & Model 3 & Forced sexual experience & 3.41 & (2.58, 4.50) \\
Class 2 vs Class 3 & Model 3 & Age, Wave III & 0.94 & (0.82, 1.07) \\
Class 2 vs Class 3 & Model 3 & Healthcare access, Wave III & 0.67 & (0.58, 0.78) \\
Class 2 vs Class 3 & Model 3 & Healthcare access, Wave IV & 1.12 & (0.98, 1.30) \\
Class 2 vs Class 3 & Model 3 & Close friends, Wave IV & 0.66 & (0.57, 0.75) \\

\midrule
Class 2 vs Class 3 & Model 4 & Forced sexual experience & 3.78 & (2.84, 5.04) \\
Class 2 vs Class 3 & Model 4 & Age, Wave III & 0.94 & (0.83, 1.08) \\
Class 2 vs Class 3 & Model 4 & Hispanic ethnicity & 1.72 & (1.07, 2.82) \\
Class 2 vs Class 3 & Model 4 & White & 1.06 & (0.48, 2.37) \\
Class 2 vs Class 3 & Model 4 & Black or African American & 2.38 & (1.08, 5.47) \\
Class 2 vs Class 3 & Model 4 & American Indian or Native American & 1.25 & (0.59, 2.67) \\
Class 2 vs Class 3 & Model 4 & Asian & 0.97 & (0.39, 2.44) \\
Class 2 vs Class 3 & Model 4 & Healthcare access, Wave III & 0.65 & (0.56, 0.76) \\
Class 2 vs Class 3 & Model 4 & Healthcare access, Wave IV & 1.09 & (0.95, 1.27) \\
Class 2 vs Class 3 & Model 4 & Close friends, Wave IV & 0.70 & (0.60, 0.80) \\

\end{longtable}
\endgroup
\end{ThreePartTable}

\section{Domain-Specific Bayesian Mixture Models}

Separate identified Bayesian mixture models were estimated for mental health burden alone and physical health burden alone (Waves III--V), using the same specification as the primary joint model.

\begin{table}[h]
\centering
\setlength{\tabcolsep}{16pt}
\renewcommand{\arraystretch}{1.15}
\begin{threeparttable}
\caption{Domain-specific Bayesian mixture models: posterior class weights.}
\small
\begin{tabular}{lcc}
\toprule
Class & Mental-health-only model & Physical-health-only model \\
\midrule
Low burden & 0.275 (0.248, 0.303) & 0.294 (0.272, 0.316) \\
Moderate burden & 0.480 (0.446, 0.514) & 0.362 (0.321, 0.399) \\
High burden & 0.245 (0.214, 0.277) & 0.344 (0.304, 0.393) \\
\bottomrule
\end{tabular}
\begin{tablenotes}[flushleft]
\footnotesize
\item \textbf{Note.} Both domain-specific models recovered a clearly ordered three-class structure with $\hat{R}=1.00$ for all key parameters. Mean maximum posterior probability was 0.858 (mental health) and 0.955 (physical health).
\end{tablenotes}
\end{threeparttable}
\end{table}

\section{Exploratory Pathway and Moderation Analyses}

Candidate pathway variables (Wave III healthcare access, Wave IV healthcare access, Wave IV closeness to friends) were evaluated using two complementary exploratory analyses, not formal causal mediation: (1) Bayesian resource models estimating the association between forced sexual experience and each candidate pathway variable, and (2) posterior predicted class probability contrasts at one SD above versus below the pathway variable's mean, among women reporting forced sexual experience.

\begin{table}[h]
\centering
\setlength{\tabcolsep}{16pt}
\renewcommand{\arraystretch}{1.15}
\begin{threeparttable}
\caption{Posterior predicted probability contrasts from Bayesian class-membership Model 3.}
\small
\begin{tabular}{p{4.4cm}p{3.3cm}cc}
\toprule
Contrast & Latent class & Mean difference & 95\% CrI \\
\midrule
High vs.\ low Wave III healthcare access & Moderate/worsening & 0.008 & (-0.070, 0.086) \\
High vs.\ low Wave III healthcare access & High chronic burden & -0.104 & (-0.187, -0.022) \\
High vs.\ low Wave III healthcare access & Low-burden/resilient & 0.096 & (0.060, 0.133) \\
High vs.\ low close friends & Moderate/worsening & 0.098 & (0.029, 0.167) \\
High vs.\ low close friends & High chronic burden & -0.152 & (-0.228, -0.077) \\
High vs.\ low close friends & Low-burden/resilient & 0.055 & (0.023, 0.087) \\
\bottomrule
\end{tabular}
\end{threeparttable}
\end{table}

Moderation analyses testing forced sexual experience $\times$ healthcare access and forced sexual experience $\times$ closeness to friends interactions found no clear evidence that these resources buffer the exposure--class-membership association specifically (e.g., forced sexual experience $\times$ Wave III healthcare access, high chronic burden vs.\ low-burden/resilient: OR = 0.75, 95\% CrI 0.53--1.03), consistent with these resources being associated with trajectory membership broadly.

\section{Exposure-Definition and Survey-Weighting Sensitivity Analyses}

\begin{table}[h]
\centering
\setlength{\tabcolsep}{10pt}
\renewcommand{\arraystretch}{1.15}
\begin{threeparttable}
\caption{Alternative forced sexual experience definitions: posterior odds ratios for high chronic burden vs.\ low-burden/resilient membership.}
\small
\begin{tabular}{lcc}
\toprule
Exposure definition & OR & 95\% CrI \\
\midrule
Any (primary definition) & 3.78 & (2.84, 5.04) \\
Physical forced sexual experience & 4.57 & (3.23, 6.58) \\
Nonphysical forced sexual experience & 3.01 & (2.27, 3.99) \\
\bottomrule
\end{tabular}
\end{threeparttable}
\end{table}

\begin{sloppypar}
A survey-weighted Bayesian class-membership model (Wave V longitudinal weight, normalized to mean 1) produced materially similar odds ratios to the primary unweighted model (forced sexual experience, high chronic burden: OR = 3.00, 95\% CrI 2.29--3.96; moderate/worsening: OR = 1.34, 95\% CrI 1.04--1.74). A survey-weight descriptive sensitivity analysis found similar weighted and unweighted class proportions (high chronic burden: 28.4\% unweighted vs.\ 30.4\% weighted) and exposure prevalence (51.7\% weighted exposure prevalence among the high chronic burden class vs.\ 25.6\% among the low-burden/resilient class).
\end{sloppypar}

\section{Prior Sensitivity for the Class-Membership Model}

\begin{sloppypar}
The fully adjusted class-membership model (Class 2, high chronic burden, vs.\ Class 3, low-burden/resilient) was re-estimated with $N(0,1^2)$, $N(0,2^2)$ (primary), and $N(0,5^2)$ priors on all logistic coefficients, using standardized predictors throughout for direct comparability across prior conditions. Table~S12 shows that posterior estimates were essentially unchanged across this threefold range in prior standard deviation, for both the primary exposure and all covariates, with $\hat{R}=1.00$--$1.00$ throughout.
\end{sloppypar}

\begin{table}[h]
\centering
\setlength{\tabcolsep}{18pt}
\renewcommand{\arraystretch}{1.15}
\begin{threeparttable}
\caption{Prior sensitivity of the fully adjusted class-membership model (Class 2 vs.\ Class 3), standardized predictors.}
\footnotesize
\begin{tabular}{p{4.2cm}>{\centering\arraybackslash}p{2.4cm}>{\centering\arraybackslash}p{2.4cm}>{\centering\arraybackslash}p{2.4cm}}
\toprule
Predictor & $N(0,1^2)$ OR (95\% CrI) & $N(0,2^2)$ OR (95\% CrI) & $N(0,5^2)$ OR (95\% CrI) \\
\midrule
Any forced sexual experience & 1.89 (1.65, 2.17) & 1.90 (1.65, 2.18) & 1.90 (1.66, 2.18) \\
Age, Wave III & 0.94 (0.82, 1.08) & 0.94 (0.82, 1.09) & 0.94 (0.83, 1.08) \\
Hispanic ethnicity & 1.17 (1.02, 1.34) & 1.17 (1.02, 1.35) & 1.17 (1.02, 1.35) \\
White & 1.05 (0.73, 1.51) & 1.07 (0.73, 1.57) & 1.06 (0.73, 1.57) \\
Black or African American & 1.49 (1.04, 2.14) & 1.52 (1.06, 2.21) & 1.52 (1.05, 2.22) \\
American Indian or Native American & 1.05 (0.91, 1.22) & 1.05 (0.91, 1.23) & 1.05 (0.90, 1.22) \\
Asian & 1.00 (0.84, 1.19) & 1.01 (0.84, 1.21) & 1.01 (0.84, 1.21) \\
Healthcare access, Wave III & 0.66 (0.56, 0.76) & 0.65 (0.56, 0.76) & 0.65 (0.56, 0.76) \\
Healthcare access, Wave IV & 1.09 (0.94, 1.27) & 1.09 (0.95, 1.27) & 1.09 (0.95, 1.27) \\
Close friends, Wave IV & 0.70 (0.60, 0.81) & 0.70 (0.60, 0.80) & 0.70 (0.60, 0.80) \\
\bottomrule
\end{tabular}
\begin{tablenotes}[flushleft]
\footnotesize
\item \textbf{Note.} CrI = credible interval. All predictors were standardized (mean-centered, unit variance) for this sensitivity analysis, so odds ratios here are per-SD and are not directly comparable to the raw-coded odds ratios reported in the main text; the purpose of this table is solely to compare estimates \emph{across} prior specifications, not to replace the primary reported odds ratios. All parameters had $\hat{R}=1.00$ across all three prior conditions.
\end{tablenotes}
\end{threeparttable}
\end{table}

\section{STROBE Checklist}

This report follows the STROBE (Strengthening the Reporting of Observational Studies in Epidemiology) guidelines for cohort studies. Table~S13 indicates where each STROBE item is addressed in the main text or Supplementary Material.

\begin{table}[h]
\centering
\setlength{\tabcolsep}{16pt}
\renewcommand{\arraystretch}{1.15}
\begin{threeparttable}
\caption{STROBE checklist for cohort studies.}
\scriptsize
\begin{tabular}{p{2.1cm}p{5.5cm}p{5.5cm}}
\toprule
Item & STROBE recommendation & Location \\
\midrule
Title/abstract & Study design and summary in title/abstract & Title; structured abstract \\
Background & Scientific background and rationale & Introduction \\
Objectives & Specific objectives/hypotheses & Introduction, final paragraph \\
Design & Study design early in the paper & Methods, Data source and sample \\
Setting & Setting, locations, dates & Methods, Data source and sample \\
Participants & Eligibility criteria, sources, methods of selection & Methods, Data source and sample; Supplementary Table~S0 \\
Variables & Outcomes, exposures, predictors, confounders & Methods, Exposure and Trajectory indicators subsections \\
Data sources/ measurement & Details of methods of assessment & Methods, Trajectory indicators subsection \\
Bias & Efforts to address bias & Methods, Statistical analysis (robustness specifications); Limitations \\
Study size & How study size was arrived at & Methods, Data source and sample \\
Quantitative variables & How variables were handled in analysis & Methods, Statistical analysis \\
Statistical methods & All statistical methods, including subgroups and sensitivity analyses & Methods, Statistical analysis; Results, Sensitivity Analyses; Supplementary Material \\
Participants (Results) & Numbers at each stage & Supplementary Table~S0 \\
Descriptive data & Characteristics of participants & Table~1 (main text) \\
Outcome data & Outcome/class data & Results \\
Main results & Estimates and precision & Results; Table~2 (main text) \\
Other analyses & Sensitivity, subgroup analyses & Results, Sensitivity Analyses; Supplementary Material \\
Key results & Summary relative to objectives & Discussion, opening paragraph \\
Limitations & Discussion of bias, imprecision & Discussion, Limitations \\
Interpretation & Cautious overall interpretation & Discussion \\
Generalizability & External validity & Discussion, Limitations \\
Funding & Funding source and role & Funding statement \\
\bottomrule
\end{tabular}
\begin{tablenotes}[flushleft]
\footnotesize
\item \textbf{Note.} This checklist reflects the standard STROBE items for cohort studies, adapted to this Bayesian secondary-data analysis; some items (e.g., loss to follow-up) are addressed via the complete-case attrition analysis (Supplementary Table~S1/S2) rather than prospective follow-up loss, given the secondary, cross-sectional-at-each-wave nature of the public-use Add Health extract used here.
\end{tablenotes}
\end{threeparttable}
\end{table}

\section{Appendix A: Exploratory Equity Analysis}

An exploratory equity analysis compared predicted class probabilities by Hispanic ethnicity using the fully adjusted class-membership model (Model 4). Predicted probability of high chronic burden membership was higher among Hispanic women relative to non-Hispanic women; predicted probability of low-burden/resilient membership was correspondingly lower. This result should be interpreted cautiously because the analysis was not designed as a fully powered intersectional equity analysis. Future work should examine structural, socioeconomic, immigration-related, linguistic, geographic, or healthcare-system factors that may contribute to unequal recovery trajectories, staging one moderator or subgroup contrast at a time given public-use sample size constraints.

\begin{sloppypar}
\section{Appendix B: Partner-Forced-Sex Wave III/IV Complete-Case Sensitivity}

A complete-case sensitivity analysis examined relationship-specific partner-forced-sex indicators from Waves III and IV (missing for 36.6\% of the analytic sample; Wave V partner-forced-sex coding was not included because its distribution suggested a skip-pattern or coding issue requiring additional codebook verification). Partner forced sex at Waves III or IV was not clearly associated with moderate/worsening burden membership (OR = 1.49, 95\% CrI 0.92--2.48) but was associated with higher posterior odds of high chronic burden membership (OR = 2.33, 95\% CrI 1.39--3.96), supporting the primary finding while highlighting the need for restricted-use or codebook-verified relationship-specific exposure measures.
\end{sloppypar}

\clearpage
\section{Supplementary Figures}

\begin{figure}[H]
\centering
\includegraphics[width=0.95\linewidth]{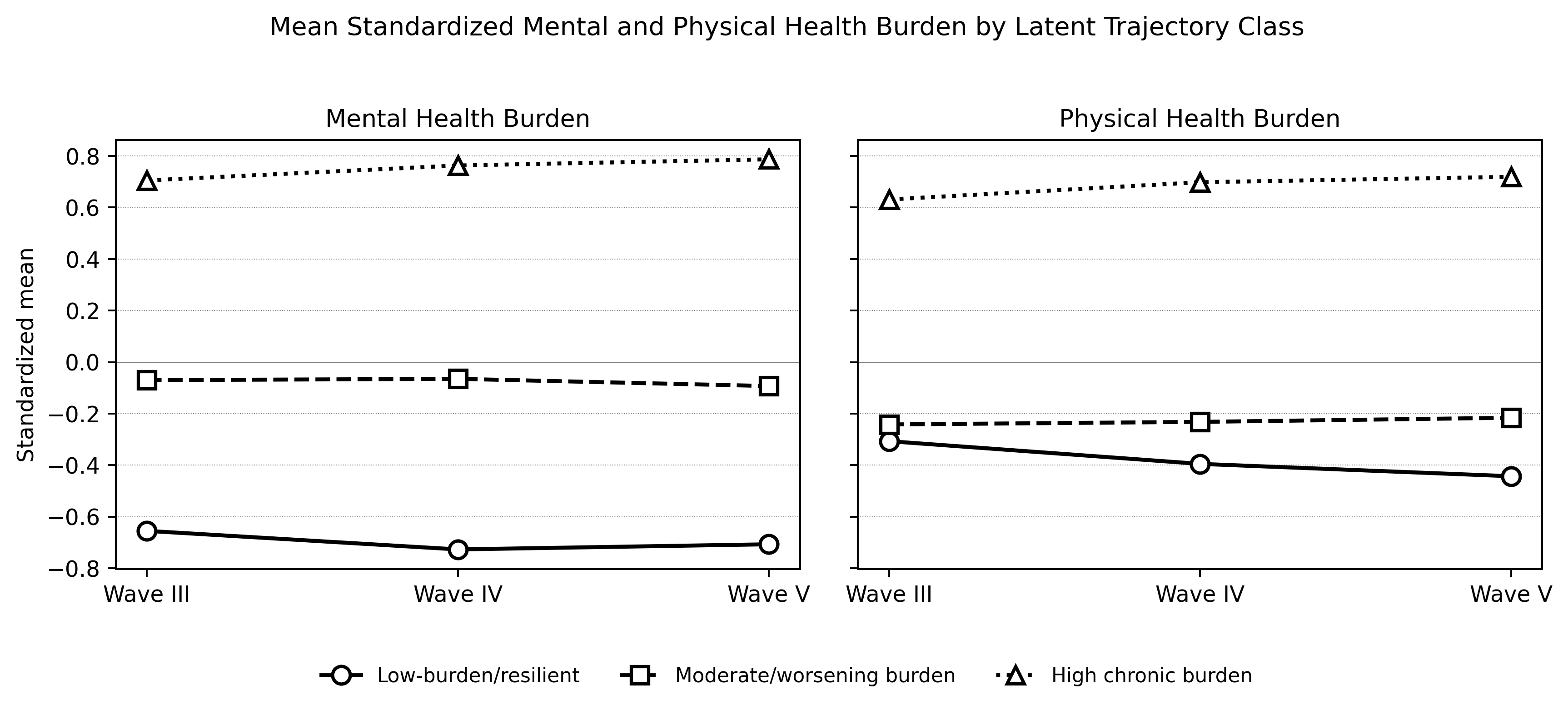}
\caption{Mean standardized mental and physical health burden by latent trajectory class, Waves III--V.}
\end{figure}

\begin{figure}[H]
\centering
\includegraphics[width=0.95\linewidth]{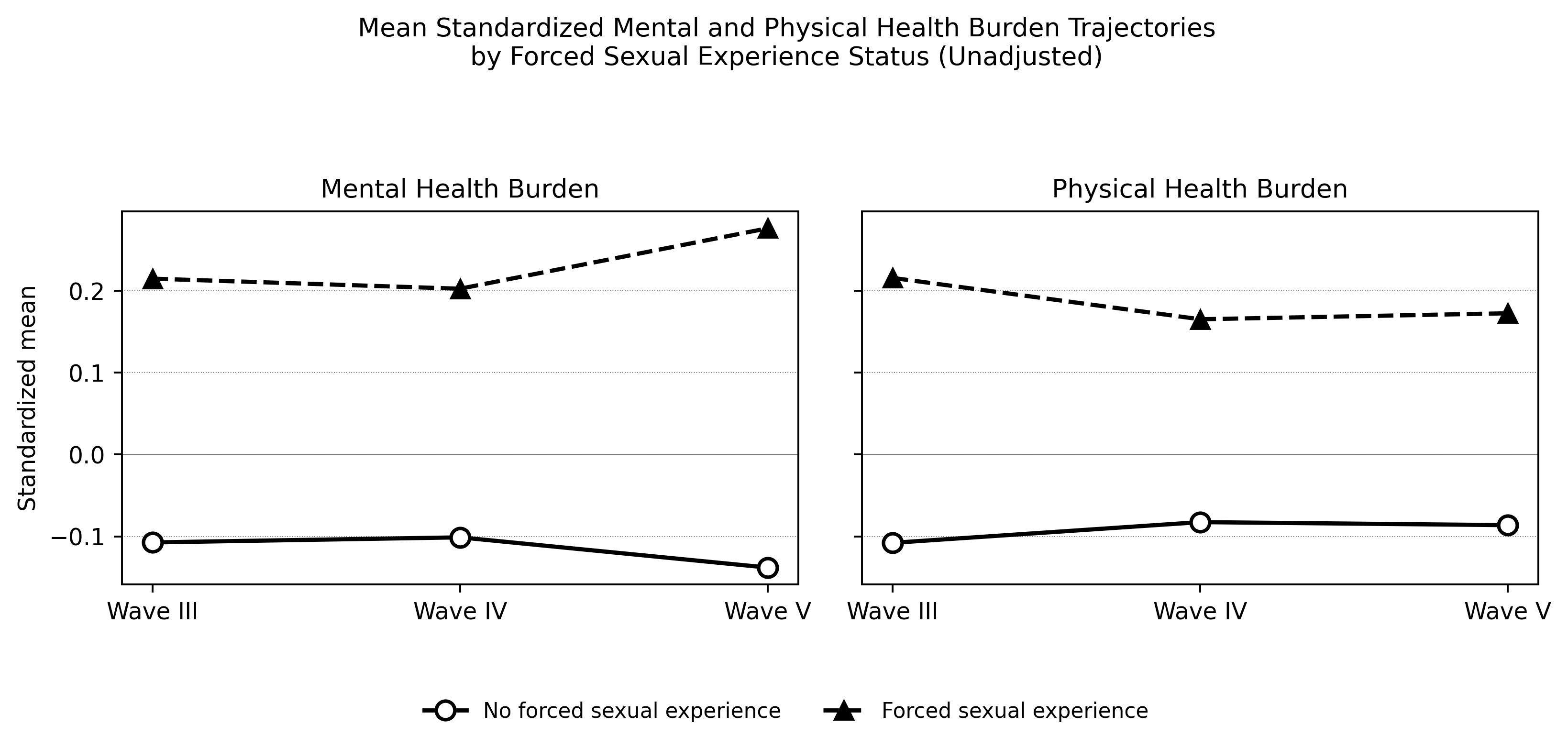}
\caption{Mean standardized mental and physical health burden trajectories by forced sexual experience status (unadjusted, pre-classification).}
\end{figure}

\begin{figure}[H]
\centering
\includegraphics[width=1.0\linewidth]{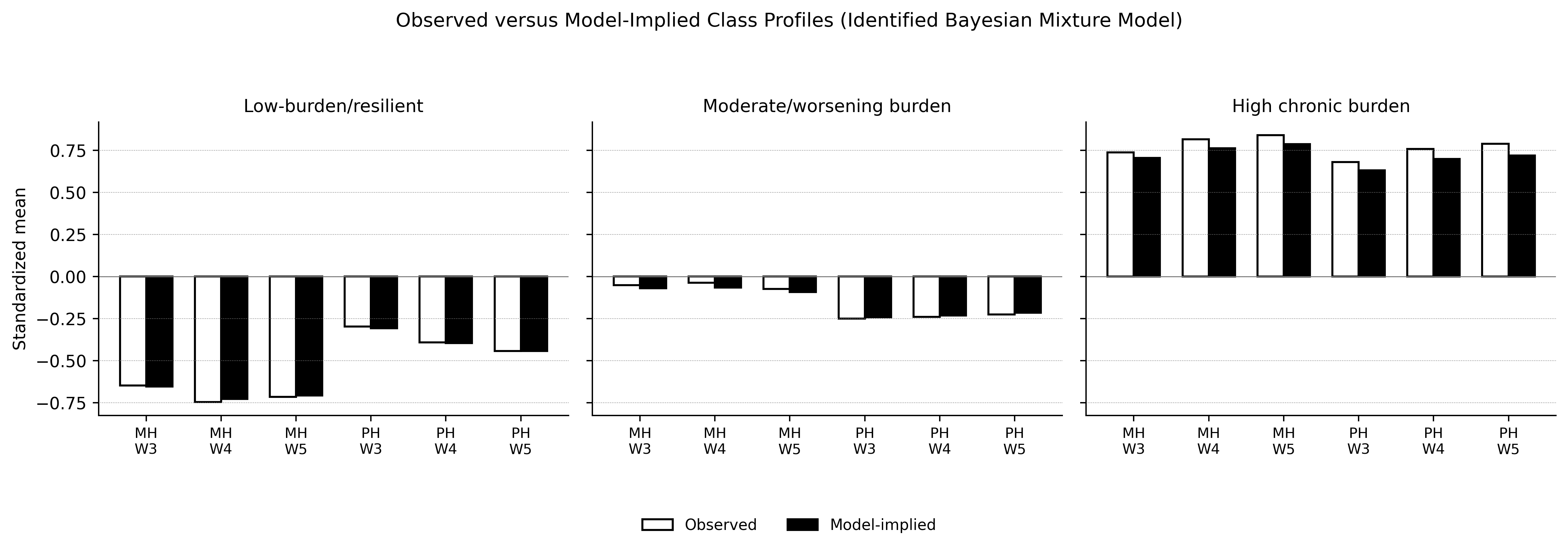}
\caption{Observed versus model-implied class profiles, identified Bayesian mixture model.}
\end{figure}

\begin{figure}[H]
\centering
\includegraphics[width=1.0\linewidth]{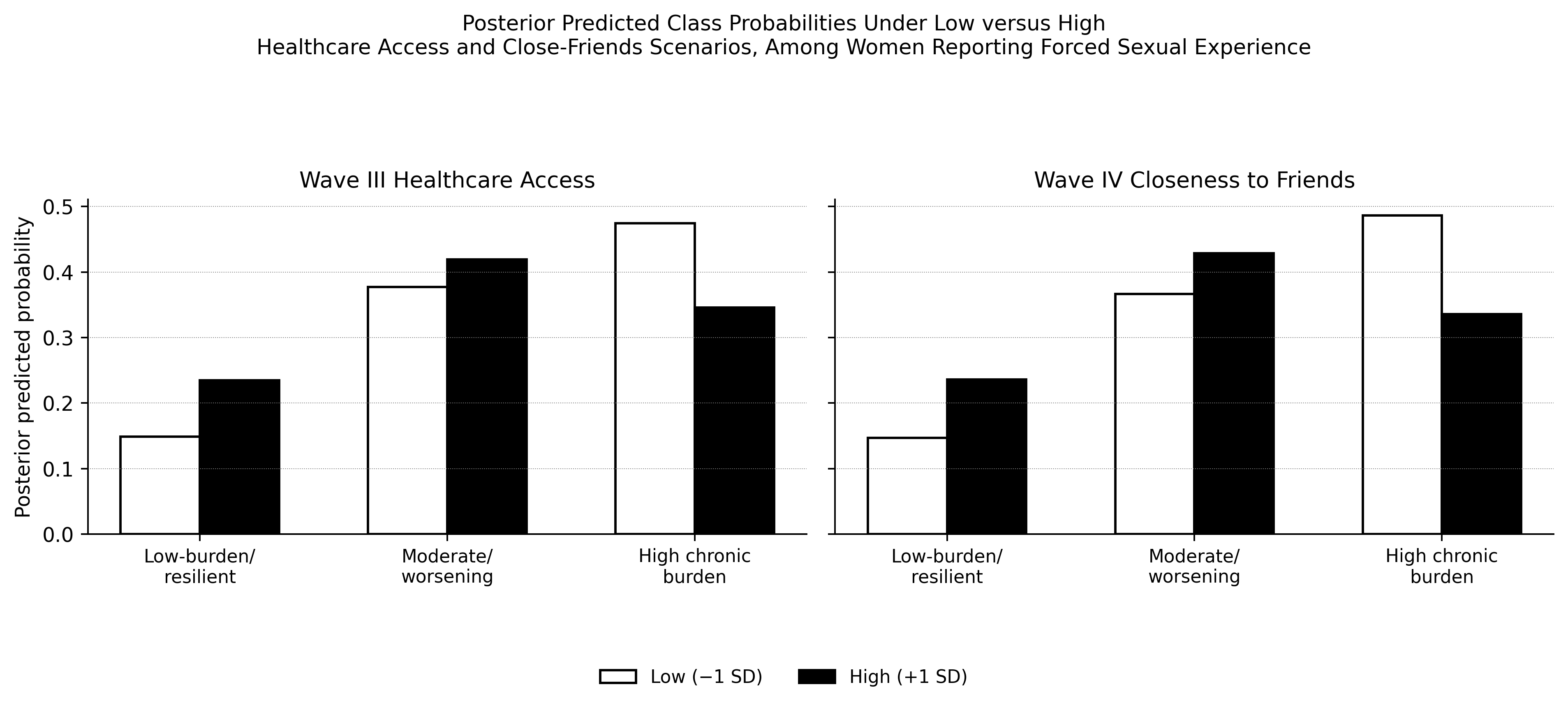}
\caption{Posterior predicted class probabilities under low versus high healthcare access and close-friends scenarios, among women reporting forced sexual experience.}
\end{figure}